\documentclass[reprint,pre,superscriptaddress]{revtex4-2}
\usepackage{graphicx}

\begin{document}


\title{Magnetic field and thermal radiation induced entropy generation in a multiphase non-isothermal plane Poiseuille flow }


\author{Joydip Chaudhuri}
\thanks{Corresponding Author}
\email{joydipchaudhuri1@gmail.com}
\affiliation{Department of Chemical Engineering, Indian Institute of Technology Guwahati, Assam 781039, India}



\date{\today}

\begin{abstract}
The effect of radiative heat transfer on the entropy generation in a two-phase non-isothermal fluid flow between two infinite horizontal parallel plates under the influence of a constant pressure gradient and transverse non-invasive magnetic field have been explored. Both the fluids are considered to be viscous, incompressible, immiscible, Newtonian and electrically conducting. The governing equations in Cartesian coordinate are solved analytically with the help of appropriate boundary conditions to obtain the velocity and temperature profile inside the channel. Application of transverse magnetic field is found to reduce the throughput and the temperature distribution of the fluids in a pressure-driven flow. The temperature and fluid flow inside the channel can also be non-invasively altered by tuning the magnetic field intensity, temperature difference between the channel walls and the fluids, and several intrinsic fluid properties. The entropy generation due to the heat transfer, magnetic field, and fluid flow irreversibilities can be controlled by altering the Hartmann number, radiation parameter, Brinkmann number, filling ratio, and the ratios of fluid viscosities, thermal and electrical conductivities. The surfaces of the channel wall are found to act as a strong source of entropy generation and heat transfer irreversibility. The rate of heat transfer at the channel walls can also be tweaked by the magnetic field intensity, temperature differences, and the fluid properties. The proposed strategies in the present study can be of significance in the design and development of gen-next microscale reactors, micro heat exchangers, and energy harvesting devices. 
\end{abstract}


\maketitle

\section{Introduction}
\indent Over the past few decades, scientific researches related to the micro or nanoscale transport phenomena have shown a pathway to miniaturization owing to the rapid development of microfluidic devices such as micro-electro-mechanical systems (MEMS) \cite{burns1998,tsuchiya2008}, drug delivery systems \cite{capretto2013}, chemical separation devices \cite{ohno2008}, polymerase chain reactors (PCR) \cite{marcus2006,zhang2012,eastburn2013}, cell sorting devices \cite{gossett2010,mazutis2013}, microreactors \cite{jensen2001,niu2009}, micromixers \cite{tice2003}, energy harvesting devices \cite{marin2013,bhattacharjee2016}, microscale biochemical analyzers \cite{taly2012}, $\mu$-total-analysis-system ($\mu$-TAS) \cite{manz1990,woolley1994}, and biomedical instruments \cite{bennet2011} to name a few. In these microfluidic flow systems, the dimensionless Reynolds number (Re) is almost always of a very small value, which defines that the flow in such systems are always laminar in nature \cite{convery2019,hudson2010}. A small value of Re is indicative of a dominant viscous force present in these flow systems. Due to these stronger viscous forces present, the laminar flow inside it can be effectively be classified as a plane Poiseuille flow if the channel walls are considered to be non-slipping and of negligible wall roughness \cite{wang2007}. However, due to the severe confinement and reduced length scales in such microfluidic devices, some unfavorable attributes have appeared in these flow systems which are driven by pressure actuation \cite{foroughi2011,timung2015,ward2005,cubaud2009,boruah2018} in general, such as friction induced power losses, biological sample dispersion, lack of precise control and high flow-rate requirement.\\
\indent In view of the aforementioned detrimental attributes, a non-invasive handle in the form of externally applied field can add more flexibility to actuate such pressure driven flows inside such micro domains. In the recent past, externally applied electric \cite{ristenpart2009,chaudhuri2017,tan2014,chaudhuri2019}, magnetic \cite{tsuchiya2008,liggieri1994,weston2010,chaudhuri2018}, photonic \cite{chraibi2008,delville2009,chaudhuri2021}, or acoustic \cite{meacham2005,friend2011} fields have been employed to disrupt the regular pressure-driven flows to improve the transport properties in such small length scales. Such external triggers help in effectively control the balance between capillary, viscous, and inertial forces in order to augment the separation of flow features \cite{mazutis2013}, improve surface to volume ratio \cite{sharma2015}, increase throughput \cite{timung2017}, and enable mixing \cite{tice2003}. Among these external field driven systems, magnetohydrodynamic (MHD) and electro-magnetohydrodynamic (EMHD) micropumps and devices in general, driven by Lorentz force, have engrossed the scientific community due to certain advantages such as, (i) absence of any mechanical moving parts, (ii) \textit{in situ} flow reversibility and (iii) higher throughput \cite{lemoff2000,bau2003,qian2009,gelb2004}. \\
\indent Design and fabrication of such a device requires an energy optimization that should ponder on the various dissipative processes existing in such microflows which directly or indirectly affect their performance and efficacy. In fact, the non-intrusive application of electromagnetic fields in these devices introduces an additional energy dissipation that, along with fluid friction and heat transfer irreversibilities, must be measured carefully in order to provide the necessary power input to perform a certain task. However, miniaturization of these microfluidic devices demands an optimal power input to perform the desired tasks in order to achieve the better overall efficiency and performance compared to the macro counterpart. In order to reduce the overall energy utilization, some efficient microfluidic systems have been designed to reduce the amount of useless energy in the recent times. Due to the impacts of energy dissipation owing to heat transfer and fluid friction, these useless energies can initiate the irreversibility of a fluidic system. In this regard, the concept of entropy generation minimization has engrossed much attention in fluid-thermal engineering and one can minimize the entropy generation by optimizing the design parameters of such systems. \\
\indent In view of this background, the detailed analysis of entropy generation seems to be an appropriate tool to assess the inherent irreversibilities in such microflows and to regulate optimized operating conditions that ensures a minimum energy dissipation consistent with the other physical constraints required by the system. Bejan introduced the analysis of entropy generation \cite{bejan1980,bejan1995} which has been employed to assess the performance of various engineering applications such as, heat exchangers \cite{san2010}, two-phase flows \cite{revellin2009}, or fuel cells \cite{sciacovelli2009}, among many others. It has also been applied to ensure the optimization of energy input and output in case of MHD flows in MHD pumps, electric generators \cite{salas1999,ibanez2002,ibanez2006,saidi2007} and fusion reactors \cite{ibanez2008}. The prior art suggests that, the analysis of entropy generation in microchannels considering mainly the effects of viscous and thermal irreversibilities \cite{haddad2004,abbassi2007} is capable of uncovering the detailed physics associated with such microflows. \\
\indent EMHD induced heat transfer and the associated entropy generation plays a major role in the field of heat and momentum transfer owing to its multifarious applications in the liquid metal flows in the metallurgical industry \cite{branover1983}, micropumps \cite{chaudhuri2018}, chemical \cite{jensen2001,niu2009}, bio-medical \cite{capretto2013} and biological \cite{gossett2010, mazutis2013} sectors. Previous studies have also attempted to decipher the temperature distribution and heat transfer characteristics of EMHD flows, where the heat generation was due to the inherent Joule heating \cite{duwairi2007} or the electro-kinetic effects associated with such flows \cite{chakraborty2013}. However, these studies have ignored the coupling of the interactions between electric and magnetic fields in the energy equation. Jian \textit{et al.} introduced these coupling terms in the energy equation to investigate the transient EMHD heat transfer and entropy generation in a parallel plate microchannel \cite{jian2015}. In the recent past, the study of entropy generation in a bilayer electroosmotic flow uncovered the velocity and temperature distribution and analytically evaluated the rate of entropy generation \cite{jian2015}. Apart from these, the entropy generation in EMHD flows and thermal transport characteristics of non-Newtonian fluids \cite{abdulhameed2017,abdulhameed2017a} and nano-fluids \cite{daniel2018} have fascinated numerous researchers in the recent past. \\
\indent The literature deliberated above discusses the entropy generation analysis for a multitude of external fields induced fluid flows in a narrow conduit. However, all of these past researches involve a single-phase fluid flow in between two infinitely long but of narrow width plates. Also, the literature discussed does not necessarily consider these type of bilayer EMHD flow under the influence of both magnetic field and radiative heat transfer. The literature also did not provide a detailed parametric analysis of the heat transfer through temperature variations and the rate of heat transfer through the walls, describing the functioning of a potential micropump under scenarios analogous to the applications involving thermal management. The literature discussed so far suggest that, a detailed analysis on the entropy generation in a multiphase non-isothermal plane Poiseuille bilayer fluid flow due to the combined effect of transverse magnetic field and thermal radiation is yet to make an appearance in the literature. More specifically, in order to utilize the potential of relatively new technologies of EMHD devices, it is of vital importance to study the associated thermal characteristics within a more generic framework consisting of a bilayer flow and by considering all the essential aspects of the microscale physics. \\
\begin{figure}
\includegraphics[width=1.0\linewidth]{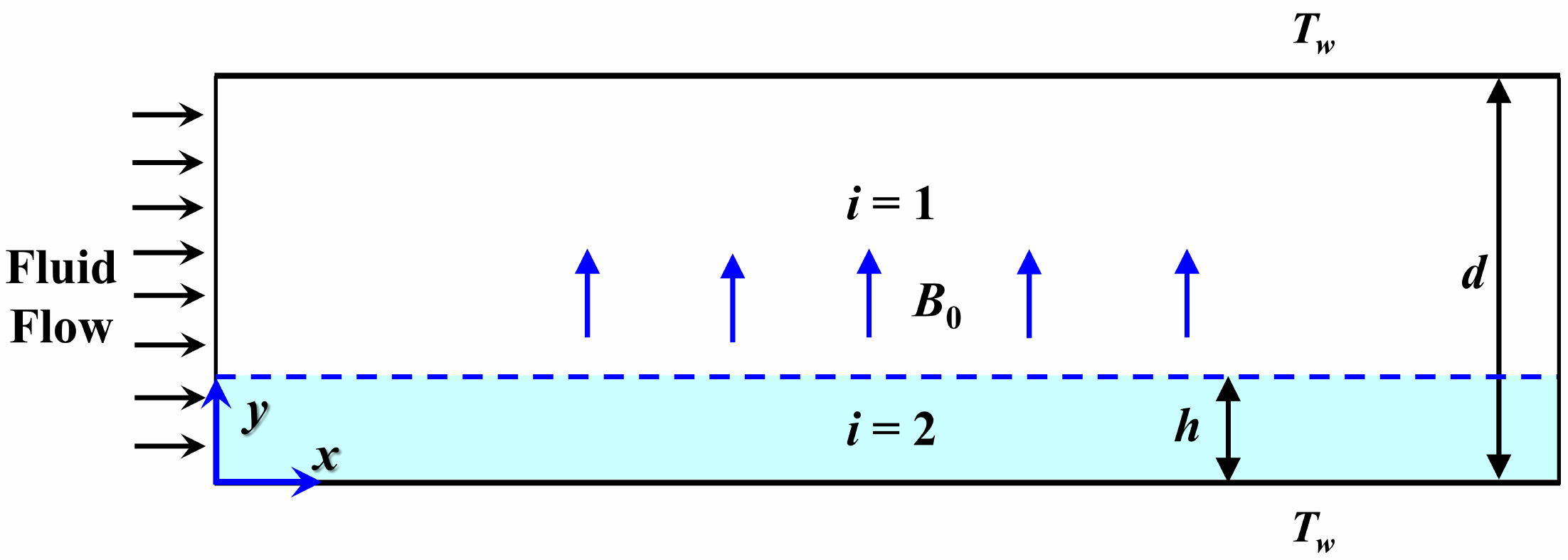}%
\caption{\label{fig1}Schematic diagram of a bilayer \textit{x}-directional fluid flow under the combined influence of a transverse uniform magnetic field and radiative heat transfer from the parallel plates (not to scale). The distance between the parallel plates is \textit{d} and the thickness of the lower layer of fluid (\textit{i} = 2, cyan region) is \textit{h}. The temperature of both the parallel plates are constant and denoted by \(T_w\). The strength of the uniform magnetic field is symbolized by \(B_0\).}
\end{figure}
\indent Herein, we analytically explore the characteristics of the entropy generation in a two-phase, bilayer, non-isothermal plane Poiseuille flow under the influence of a non-invasively applied transverse magnetic field and considering the effect of radiative heat transfer. The schematic diagram in figure 1 shows the typical geometry chosen for the analytical model. Employing fully developed, steady, laminar Poiseuille flow model on a pair of electrically conducting, incompressible, non-isothermal, and immiscible fluids under the influence of a magnetic field and radiative heat transfer, an exact solution of governing equations for both the fluid regions has been obtained in closed form. Contributions to the total entropy generation in the system due to heat transfer, magnetic field and radiation are also evaluated analytically. The velocity and temperature distribution inside the channel and entropy generation are found to be dependent on various fluid properties such as, viscosity, thermal and electrical conductivities, as well as, on a number of external field parameters namely, magnetic field intensity, conductive and the radiative heat transfer parameters. A detailed parametric study has been carried out to see the effect of these pertinent parameters on the flow field, temperature field and the entropy generation characteristics. We have also uncovered the contribution of heat transfer, magnetic field, and fluid flow irreversibilities to the total entropy generation for such a system, which provides an idea about the input energy budget for such a flow.  \\
\indent The remainder of the paper is organized in the following manner. In section II, the details of the theory and the problem formulation are discussed. It also contains the details of the dimensional and non-dimensional governing equations, boundary conditions, expressions for the distribution of velocity, temperature, and the entropy generation. The results are discussed in detail in section III, before the conclusions are drawn in section IV.  \\
\section{Theoretical Formulation}
\subsection{Problem formulation}
We consider fully developed, steady, laminar Poiseuille flow of a pair of electrically conducting, incompressible and immiscible fluids (\(i\) = 1 and 2) in a channel bounded by two plates under the effect of a transverse magnetic field \(B_0\), which is applied normal to the flow direction. Both the parallel plates are impermeable and maintained at a uniform temperature \(T_w\). The \textit{x}-axis is considered to be in the direction of fluid flow along the length of the channel, whereas, the \textit{y}-axis is taken normal to the lower plate along the width of the channel. The parallel plates are considered to be infinitely long and the distance between them is \textit{d}. The thickness of the lower layer of fluid (\textit{i} = 2) is \textit{h}. A representative schematic diagram of the problem under consideration is shown in figure 1. For the \textit{i}\textsuperscript{th} fluid, the notations \({\mathbf{u}}_i\left(u_i,v_i\right)\), \(t_i\), \(\rho_i\), \(\eta_i\), \(\sigma_i\), and \(k_i\) denote the velocity vector (velocity components), temperature, density, viscosity, electrical conductivity, and thermal conductivity, respectively. The magnetic Reynolds number is assumed to be small, so that the induced magnetic field can be neglected and the Hall-effect of MHD is assumed to be negligible.
\subsection{Dimensional governing equations and boundary conditions}
The dynamics of the steady state fully developed laminar Poiseuille flow of both the fluids (\textit{i} = 1 and 2) in a channel is governed by the equations of motions and continuity equation, which can be expressed as,
\begin{equation}\label{1}
-{{\mathbf{\nabla}}p_i}+{\mathbf{\nabla}}\cdot\left[\eta_i\left({\mathbf{\nabla}}{\mathbf{u}_i}+{\mathbf{\nabla}}{\mathbf{u}_i^{T}}\right)\right]+{\mathbf{F}_M}=0,
\end{equation}
\begin{equation}\label{2}
{\mathbf{\nabla}}\cdot{\mathbf{u}_i}=0.
\end{equation}
\indent The additional body force (\({\mathbf{F}_M}\)) in equation (\ref{1}), also termed as Lorentz force, is added due to the imposed uniform magnetic field (\(\mathbf{B}_0\)) along the \(y\)-direction. Also, it is assumed that the associated electric field due to the application of uniform magnetic field is zero, i.e., \(\mathit{\mathbf{E}}_0=\nabla{V}=0\). Therefore, from Ohm’s law, we can express the current density (\(\mathbf{J}\)) as, \(\displaystyle \mathbf{J}_i=\sigma_i\left(-\nabla{V}_i+{\mathbf{u}_i}\times{\mathbf{B}_0}\right)=\sigma_i\left({\mathbf{u}_i}\times{\mathbf{B}_0}\right)\). The Lorentz force (\({\mathbf{F}_M}\)) in equation (\ref{1}) then can be evaluated as, \(\displaystyle {\mathbf{F}_M}=\mathbf{J}_i\times{\mathbf{B}_0}=\sigma_i\left({\mathbf{u}_i}\times{\mathbf{B}_0}\right)\times{\mathbf{B}_0}=-\sigma_i{u_i}B_0^2{\mathbf{\hat{x}}}\), where, \(\mathbf{\hat{x}}\) is the unit vector along the \(x\)-direction. Since, the fluid flow in the channel is considered to be only \(x\)-directional, therefore, applying the assumptions, the dimensional governing equations for both the fluids can be expressed as follows.\\
\indent For \textit{i} = 1, (\(h\leq{y}\leq{d}\)),
\textit{x}-momentum balance equation can be written as, 
\begin{equation}\label{3}
\eta_1\frac{d^2{u_1}}{dy^2}-\sigma_1{u_1}B_0^2=\frac{dp}{dx},
\end{equation}
\textit{y}-momentum balance equation can be written as, 
\begin{equation}\label{4}
\frac{dp}{dy}=0,
\end{equation}
the continuity equation can be expressed as, 
\begin{equation}\label{5}
\frac{du_1}{dx}=0,
\end{equation}
and the energy balance equation can be expressed as, 
\begin{equation}\label{6}
k_1\frac{d^2{t_1}}{dy^2}+\eta_1\left(\frac{du_1}{dy}\right)^2+\sigma_1{u_1^2}B_0^2-\frac{dq_1^r}{dy}=0.
\end{equation}
\indent For \textit{i} = 2, (\(0\leq{y}\leq{h}\)),
\textit{x}-momentum balance equation can be written as, 
\begin{equation}\label{7}
\eta_2\frac{d^2{u_2}}{dy^2}-\sigma_2{u_2}B_0^2=\frac{dp}{dx},
\end{equation}
\textit{y}-momentum balance equation can be written as, 
\begin{equation}\label{8}
\frac{dp}{dy}=0,
\end{equation}
the continuity equation can be expressed as, 
\begin{equation}\label{9}
\frac{du_2}{dx}=0,
\end{equation}
and the energy balance equation can be expressed as, 
\begin{equation}\label{10}
k_2\frac{d^2{t_2}}{dy^2}+\eta_2\left(\frac{du_2}{dy}\right)^2+\sigma_2{u_2^2}B_0^2-\frac{dq_2^r}{dy}=0.
\end{equation}
\indent The \(y\)-momentum balance and the continuity equation for both the fluids essentially signify that, \(p\neq{f\left(y\right)}\) and \(u_i\neq{f\left(x\right)}\). \\
\indent The dimensional boundary conditions can be stated as,
\begin{equation}
\mathrm{at}\hspace{2mm}y=0,\hspace{2mm}u_2=0,\hspace{2mm}t_2=T_w,\label{11}
\end{equation}
\begin{equation}
\mathrm{at}\hspace{2mm}y=h,\hspace{2mm}u_1=u_2,\hspace{2mm}\eta_2\left(\frac{du_2}{dy}\right)=\eta_1\left(\frac{du_1}{dy}\right),\label{12}
\end{equation}
\begin{equation}\label{13}
\mathrm{at}\hspace{2mm}y=h,\hspace{2mm}t_1=t_2,\hspace{2mm}k_2\left(\frac{dt_2}{dy}\right)=k_1\left(\frac{dt_1}{dy}\right),
\end{equation}
\begin{equation}
\mathrm{and\hspace{2mm}at}\hspace{2mm}y=d,\hspace{2mm}u_1=0,\hspace{2mm}t_1=T_w.\label{14}
\end{equation}
\indent Here, \(B_0\) is the magnetic field intensity, \(dq_i^r\) is the radiative heat flux, and \(\displaystyle \frac{dp}{dx}\) is the applied pressure gradient. Furthermore, we neglected the effect of interface curvature at the fluid-fluid boundary in order to neglect the effect of interfacial tension and the additional Maxwell stresses along the interface \cite{reddy2011}. The no slip boundary conditions for the fluid flow near the wall mentioned in the equations (11) and (14) of the mathematical formulation are consistent with the typical microchannel flows or ultra-thin film flows \cite{khanna1997,sharma2014,sharma2015,chaudhuri2017,chaudhuri2018,chaudhuri2019}.\\
\indent Following the equilibrium model of Cogley \textit{et al.} \cite{cogley1968}, the expression for the radiative heat flux (\(dq_i^r\)) in equations (6) and (10) can be expressed as,
\begin{equation}\label{15}
\frac{dq_i^r}{dy}=4\left(t_i-T_w\right)\int_0^{\infty}K_{\lambda{w}}\left(\frac{\partial {e_{b\lambda}}}{\partial T}\right)_{w}d{\lambda}=4I^{\ast}\left(t_i-T_w\right),
\end{equation}
where, \(\displaystyle I^{\ast}=\int_0^{\infty}K_{\lambda{w}}\left(\frac{\partial {e_{b\lambda}}}{\partial T}\right)_{w}d{\lambda}\), \(K_{\lambda{w}}\) is the absorption coefficient at the plate and \({e_{b\lambda}}\) is the Planck’s constant. 
\subsection{Non-dimensional governing equations and boundary conditions}\label{C}
The equations are reduced to dimensionless forms by using the following scheme,
\begin{equation}\label{16}
\left. \begin{array}{l}
\displaystyle
\bar{u}_1=\frac{u_1}{u_0},\bar{u}_2=\frac{u_2}{u_0},\bar{y}=\frac{y}{d},{a}=\frac{h}{d},\bar{\eta}=\frac{\eta_2}{\eta_1}, \bar{k}=\frac{k_2}{k_1}, \\[10pt]
\displaystyle
\bar{\sigma}=\frac{\sigma_2}{\sigma_1}, {\theta_1}=\frac{t_1-T_w}{T_w},{\theta_2}=\frac{t_2-T_w}{T_w}, \mathrm{Ha}_1^2=\frac{\sigma_1{B_0^2}d^2}{\eta_1}, 
 \\[10pt]
\displaystyle
\mathrm{Br}_1=\frac{\eta_1{u_0^2}}{k_1T_w},\mathrm{Ha}_2^2=\frac{\sigma_2{B_0^2}d^2}{\eta_2}=\mathrm{Ha}_1^2\left(\frac{ \bar{\sigma}}{\bar{\eta}}\right),\\[10pt]
\displaystyle
\mathrm{Br}_2=\frac{\eta_2{u_0^2}}{k_2T_w}=\mathrm{Br}_1\left(\frac{\bar{\eta}}{\bar{k}}\right), F_1=\frac{4I^{\ast}d^2}{k_1}, \hspace{2mm} \mathrm{and}\\[10pt]
\displaystyle
F_2=\frac{4I^{\ast}d^2}{k_2}=F_1\left(\frac{1}{\bar{k}}\right).
\end{array} \right\}
\end{equation}\\
\indent Here, \(u_0=\displaystyle -\frac{d^2}{\eta_1}\frac{dp}{dx}\) is the reference velocity, \(a\) is the filling ratio, \(\bar{\eta}\) is the viscosity ratio, \(\bar{k}\) is the thermal conductivity ratio, \(\bar{\sigma}\) is the electrical conductivity ratio, \(\mathrm{Ha}_i\) \(\left(i=1,2\right)\) are the Hartmann numbers, \(\mathrm{Br}_i\) \(\left(i=1,2\right)\) are the Brinkman numbers, and \(F_i\) \(\left(i=1,2\right)\) are the radiation parameters. The Hartman numbers (\(\mathrm{Ha}_i\)) indicate the strength of the Lorentz forces compared to the viscous forces; the Brinkman numbers (\(\mathrm{Br}_i\)) signify the ratio between heat produced by viscous dissipation and heat transported by molecular conduction; and the radiation parameters (\(F_i\)) measure the relative dominance of radiative to conductive heat transfer within the system. \\
\indent Employing the non-dimensional parameters, the dimensionless governing equations for fluid flow and temperature distribution can be expressed as follows.\\
\indent For \textit{i} = 1, (\(a\leq{\bar{y}}\leq{1}\)),
\textit{x}-momentum balance equation can be written as, 
\begin{equation}\label{17}
\frac{d^2{\bar{u}_1}}{d\bar{y}^2}-\mathrm{Ha}_1^2{\bar{u}_1}+1=0,
\end{equation}
and the energy balance equation can be expressed as, 
\begin{equation}\label{18}
\frac{d^2{\theta_1}}{d\bar{y}^2}+\mathrm{Br}_1\left(\frac{d\bar{u}_1}{d\bar{y}}\right)^2+\mathrm{Ha}_1^2\mathrm{Br}_1{\bar{u}_1^2}-F_1\theta_1=0.
\end{equation}
\indent For \textit{i} = 2, (\(0\leq{\bar{y}}\leq{a}\)),
\textit{x}-momentum balance equation can be written as, 
\begin{equation}\label{19}
\frac{d^2{\bar{u}_2}}{d\bar{y}^2}-\frac{\mathrm{Ha}_1^2}{\bar{\eta}}\bar{u}_2+\frac{1}{\bar{\eta}}=0,
\end{equation}
and the energy balance equation can be expressed as, 
\begin{equation}\label{20}
\frac{d^2{\theta_2}}{d\bar{y}^2}+\mathrm{Br}_2\left(\frac{d\bar{u}_2}{d\bar{y}}\right)^2+\mathrm{Ha}_2^2\mathrm{Br}_2{\bar{u}_2^2}-F_2\theta_2=0.
\end{equation}
\indent The corresponding dimensionless boundary conditions can be expressed as,
\begin{equation}
\mathrm{at}\hspace{2mm}\bar{y}=0,\hspace{2mm}\bar{u}_2=0,\hspace{2mm}\theta_2=0,\label{21}
\end{equation}
\begin{equation}
\mathrm{at}\hspace{2mm}\bar{y}=a,\hspace{2mm}\bar{u}_1=\bar{u}_2,\hspace{2mm}\bar{\eta}\left(\frac{d\bar{u}_2}{d\bar{y}}\right)=\left(\frac{d\bar{u}_1}{d\bar{y}}\right),\label{22}
\end{equation}
\begin{equation}\label{23}
\mathrm{at}\hspace{2mm}\bar{y}=a,\hspace{2mm}\theta_1=\theta_2,\hspace{2mm}\bar{k}\left(\frac{d\theta_2}{d\bar{y}}\right)=\left(\frac{d\theta_1}{d\bar{y}}\right),
\end{equation}
\begin{equation}
\mathrm{and\hspace{2mm}at}\hspace{2mm}\bar{y}=1,\hspace{2mm}\bar{u}_1=0,\hspace{2mm}\theta_1=0.\label{24}
\end{equation}
\indent The mathematical description of the system is non-dimensionalized in the present study so that the applicability of the model becomes independent of the length scale of the problem. This in particular allows the non-dimensional mathematical model to be well poised to describe the momentum and heat transfer mechanisms and the associated entropy generation characteristics of a low Re microchannel flow in case the channel walls are non-slipping and of negligible roughness \cite{wang2007}. Therefore, in a sense, dimensionless description of the plane Poiseuille flow presented in this study can also be employed to properly describe the physics of a typical microchannel flow if the wall properties remain same \cite{teo2009}.
\subsection{Expressions for velocity and temperature distribution}
The non-dimensional governing equations (\ref{17}) $–$ (\ref{20}) are solved employing the corresponding boundary conditions (\ref{21}) $–$ (\ref{24}), and after dropping the overbars for convenience, the solved expressions for velocity (\(u_i\)) profiles can be given by,
\begin{equation}\label{25}
u_1=A_1\cosh\left(\mathrm{Ha}_1{y}\right)+A_2\sinh\left(\mathrm{Ha}_1{y}\right)+\frac{1}{\mathrm{Ha}_1^2},
\end{equation}
\begin{equation}\label{26}
u_2=A_3\cosh\left(M{y}\right)+A_4\sinh\left(M{y}\right)+\frac{1}{{\eta}M^2}.
\end{equation}
\indent The coefficients in equation (\ref{25}) and (\ref{26}) are given by,
\begin{equation}\label{27}
\left. \begin{array}{l}
\displaystyle
\mathrm{Ha}_1^2=\eta{M^2}, A_1=-\frac{A_2\mathrm{Ha}_1^2\sinh\left(\mathrm{Ha}_1\right)+1}{\mathrm{Ha}_1^2\cosh\left(\mathrm{Ha}_1\right)}, \\[10pt]
\displaystyle
A_2=\frac{\left(1/\mathrm{Ha}_1\right)\alpha+\left(1/M\right)\beta}{\eta{M}\gamma+\mathrm{Ha}_1\delta}, A_3=-\frac{1}{\eta{M^2}}, 
 \\[10pt]
\displaystyle
A_4=\frac{A_1\lambda+A_2\phi+\left({1}/{\mathrm{Ha}_1^2}\right)+A_3\left(1-\psi\right)}{\omega},\\[10pt]
\displaystyle
\alpha=\omega\phi,
\beta=\cosh\left(\mathrm{Ha}_1\right)-\lambda\psi, \\[10pt]
\displaystyle
\gamma=\psi\sinh\big\{\mathrm{Ha}_1\left(1-a\right)\big\}, \\[10pt]
\displaystyle
\delta=\omega\cosh\big\{\mathrm{Ha}_1\left(1-a\right)\big\}, \\[10pt]
\displaystyle
\lambda=\cosh\left(\mathrm{Ha}_1a\right),\phi=\sinh\left(\mathrm{Ha}_1a\right),\\[10pt]
\displaystyle
\psi=\cosh\left(Ma\right),\hspace{2mm} \mathrm{and}\hspace{2mm}\omega=\sinh\left(Ma\right).
\end{array} \right\}
\end{equation}\\
\indent The solved expressions for the temperature (\(\theta_i\)) profiles can be expressed as,
\begin{equation}\label{28}
\theta_1=P\left(\cosh{R}-\sinh{R}\right),
\end{equation}
\begin{equation}\label{29}
\theta_2=Q\left(\cosh{S}-\sinh{S}\right),
\end{equation}
where, \(R=\displaystyle 2y\left(\sqrt{F_1}+\mathrm{Ha}_1\right)\) and \(S=\displaystyle 2y\left(\sqrt{F_2}+M\right)\).\\
\indent The complex expressions of the coefficients \(P\) and \(Q\) in equations (\ref{28}) and (\ref{29}) are calculated using the commercial package Mathematica\textsuperscript{\textregistered}. The mathematical expressions of these coefficients are given in appendix A.
\subsection{Entropy generation}
The thermodynamic irreversibility within the system considered can be characterized by the existence of entropy in the system. Therefore, the calculation of entropy generation can be a convenient tool to quantitatively measure of irreversibility associated with the process. The local volumetric rate of entropy generation [\(\left(S^{'''}\right)_i\hspace{1mm}, \forall{i=1,2}\)] in the presence of a magnetic field for both fluids can be calculated as follows \cite{bejan1982,woods1975}.\\
\indent For \(i=1\) (\(h\leq{y}\leq{d}\)), 
\begin{equation}\label{30}
\left(S^{'''}\right)_1=\frac{k_1}{T_w}\left(\frac{dt_1}{dy}\right)^2+\frac{1}{T_w}\left[\eta_1\left(\frac{du_1}{dy}\right)^2+\sigma_1{B_0^2}{u_1^2}\right],
\end{equation}
\indent and for \(i=2\) (\(0\leq{y}\leq{h}\)), 
\begin{equation}\label{31}
\left(S^{'''}\right)_2=\frac{k_2}{T_w}\left(\frac{dt_2}{dy}\right)^2+\frac{1}{T_w}\left[\eta_2\left(\frac{du_2}{dy}\right)^2+\sigma_2{B_0^2}{u_2^2}\right].
\end{equation}
\indent The non-dimensional form of the entropy generation number (\(S_i^N\)) in both the regions can be expressed as,
\indent for \(i=1\) (\(a\leq{y}\leq{1}\)), 
\begin{equation}\label{32}
S_1^N=\frac{\left(S^{'''}\right)_1}{\left(S^{'''}\right)_0}=\left(\frac{d\theta_1}{dy}\right)^2+\mathrm{Br}_1\left[\left(\frac{du_1}{dy}\right)^2+\mathrm{Ha}_1^2{u_1^2}\right],
\end{equation}
\indent and for \(i=2\) (\(0\leq{y}\leq{a}\)), 
\begin{equation}\label{33}
S_2^N=\frac{\left(S^{'''}\right)_2}{\left(S^{'''}\right)_0}=k\left(\frac{d\theta_2}{dy}\right)^2+k\mathrm{Br}_2\left[\left(\frac{du_2}{dy}\right)^2+\mathrm{Ha}_2^2{u_2^2}\right].
\end{equation}
\begin{figure*}[t]
\includegraphics[width=0.9\linewidth]{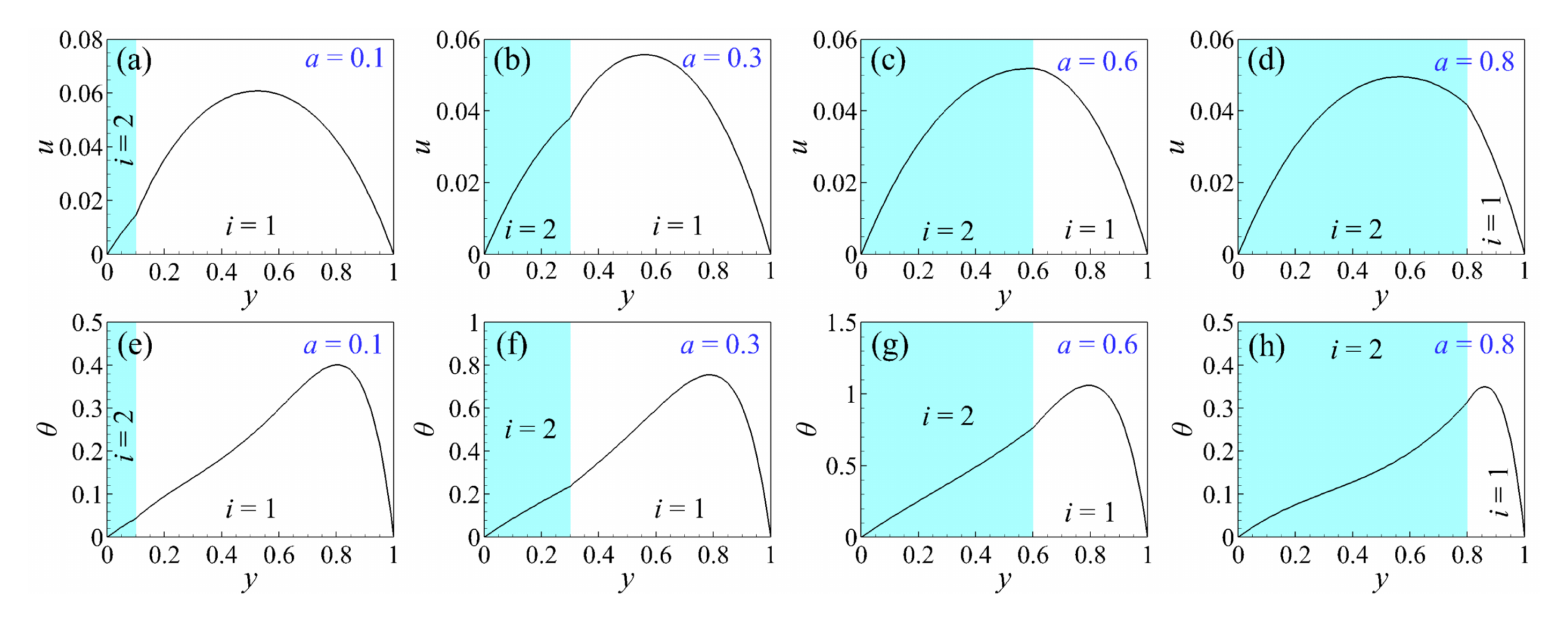}%
\caption{\label{fig2}Plots (a) $–$ (d) and (e) $–$ (h) show the velocity (\textit{u}) and temperature ($\theta$) distribution across the width (y) of the channel for different filling ratios (a), respectively. The other parameters are, \(\eta=2.0\), \(\mathrm{Ha}_1=3.0\), \(k=1.5\), \(\sigma=1.0\), \(\mathrm{Br}_1=2.0\), and \(F_1=2.0\).}
\end{figure*}
\indent In the expressions (\ref{32}) and (\ref{33}), \(\left(S^{'''}\right)_0=\displaystyle \frac{k_1}{d^2}\) is the reference volumetric entropy generation, the heat transfer irreversibility (HTI) in the fluid 1 (\(i = 1\)) is, \(\mathrm{HTI}_1=\displaystyle \left(\frac{d\theta_1}{dy}\right)^2\), the fluid friction irreversibility (FFI) in the fluid 1 (\(i = 1\)) is, \(\mathrm{FFI}_1=\displaystyle \mathrm{Br}_1\left(\frac{du_1}{dy}\right)^2\), the magnetic field irreversibility (MFI) in the fluid 1 (\(i = 1\)) is, \(\mathrm{MFI}_1=\displaystyle \mathrm{Br}_1\mathrm{Ha}_1^2{u_1^2}\), the heat transfer irreversibility (HTI) in the fluid 2 (\(i = 2\)) is, \(\mathrm{HTI}_2=\displaystyle k\left(\frac{d\theta_2}{dy}\right)^2\), the fluid friction irreversibility in the fluid 2 (\(i = 2\)) is, \(\mathrm{FFI}_2=\displaystyle k\mathrm{Br}_2\left(\frac{du_2}{dy}\right)^2\), and the magnetic field irreversibility in the fluid 2 (\(i = 2\)) is, \(\mathrm{MFI}_2=\displaystyle k\mathrm{Br}_2\mathrm{Ha}_2^2{u_2^2}\). These irreversibilities can be evaluated by using the following dimensionless numbers, termed as, Bejan numbers (\(\mathrm{Be}_i, i = 1,2\)), magnetic field irreversibility parameters (\(\mathrm{I}_i, i = 1,2\)), and fluid flow irreversibility parameters (\(\mathrm{J}_i, i = 1,2\)). The mathematical expressions of these dimensionless numbers (\(\mathrm{Be}_i, \mathrm{I}_i,\) and \(\mathrm{J}_i\)) can be given by,  
\begin{equation}\label{34}
\mathrm{Be}_i=\frac{\mathrm{HTI}_i}{S_i^N}=\frac{\left({d\theta_i}/{dy}\right)^2}{\left({d\theta_i}/{dy}\right)^2+\mathrm{Br}_i\left[\left({du_i}/{dy}\right)^2+\mathrm{Ha}_i^2{u_i^2}\right]},
\end{equation}
\begin{equation}\label{35}
\mathrm{I}_i=\frac{\mathrm{MFI}_i}{S_i^N}=\frac{\mathrm{Ha}_i^2\mathrm{Br}_i{u_i^2}}{\left({d\theta_i}/{dy}\right)^2+\mathrm{Br}_i\left[\left({du_i}/{dy}\right)^2+\mathrm{Ha}_i^2{u_i^2}\right]},
\end{equation}
\begin{equation}\label{36}
\mathrm{J}_i=\frac{\mathrm{FFI}_i}{S_i^N}=\frac{\mathrm{Br}_i\left({du_i}/{dy}\right)^2}{\left({d\theta_i}/{dy}\right)^2+\mathrm{Br}_i\left[\left({du_i}/{dy}\right)^2+\mathrm{Ha}_i^2{u_i^2}\right]}.
\end{equation}
\section{Results and Discussion}
The study of fully developed, steady, laminar Poiseuille flow of a pair of electrically conducting, incompressible fluids (\textit{i} =1 and 2) in a channel bounded by two plates under the effect of a transverse magnetic field and non-isothermal temperature field can be entirely characterized by the six fundamental physical properties, namely, the viscosities (\(\eta_i\)), thermal (\(k_i\)) and electrical (\(\sigma_i\)) conductivities. Since, we are considering the fluid flow to be in Stokes flow regime the effect of the fluid densities (\(\rho_i\)) are negligible. Apart from these, the dynamics of the problem depends on the filling ratio (\(a\)), applied magnetic field intensity (\(B_0\)) and the temperature differences (\(\theta_i\)) of both the fluids and the channel walls. Since all these ten (10) parameters involve three fundamental units (mass, time and length), the problem may be characterized with seven (7) independent dimensionless parameters. We can select these independent dimensionless parameters as, \(a\), \(\eta\), \(k\), \(\sigma\), Ha\textsubscript{1}, Br\textsubscript{1}, and \(F_1\). The physical significances of these dimensionless quantities are already described in subsection \ref{C}. \\
\indent It may be noted here that, the dimensionless numbers  Ha\textsubscript{2}, Br\textsubscript{2}, and \(F_2\) for fluid 2 can be expressed in terms of  Ha\textsubscript{1}, Br\textsubscript{1}, and \(F_1\) as shown in equation (\ref{16}). To understand the effect of these seven dimensionless parameters on the fluid flow or the velocity distribution, temperature distribution and by extension, the entropy generation within the channel, a systematic study has been carried out over a range of \(a\), \(\eta\), \(k\), \(\sigma\), Ha\textsubscript{1}, Br\textsubscript{1}, and \(F_1\).
\subsection{Effect of filling ratio (\textit{a})}
Figures 2(a) $–$ (d) show the variation of velocity (\textit{u}) along the width (\textit{y}) of the channel with increasing filling ratios (\textit{a}).  The plots (a) $–$ (d) show that with increase in \textit{a} from 0.1 to 0.8, the maximum as well as average velocity of the fluids inside the channel decreases. The viscosity ratio (\(\eta\)) in these cases is kept constant at 2. With the increase in \textit{a}, the channel gets filled with the more viscous fluid 2, which ensures that the fluid 2 as well as fluid 1 (because of the laminar nature of the flow) flows slowly inside the channel. \\
\indent Figures 2(e) $–$ (h) show the variation of temperature (\(\theta\)) along the width (\textit{y}) of the channel with increasing filling ratios (\textit{a}). The plots (e) $–$ (g) show that with increase in \textit{a} from 0.1 to 0.6, the maximum as well as average temperature of the fluids inside the channel increases. The slowly flowing fluids due to the increase in filling ratio retains more heat with them which in turn increase the temperature distribution inside the channel. However, with further increase in filling ratio (\textit{a} = 0.8), the temperature of the fluids decreases considerably as shown in figure 2(h). \\
\begin{figure*}[t]
\includegraphics[width=0.9\linewidth]{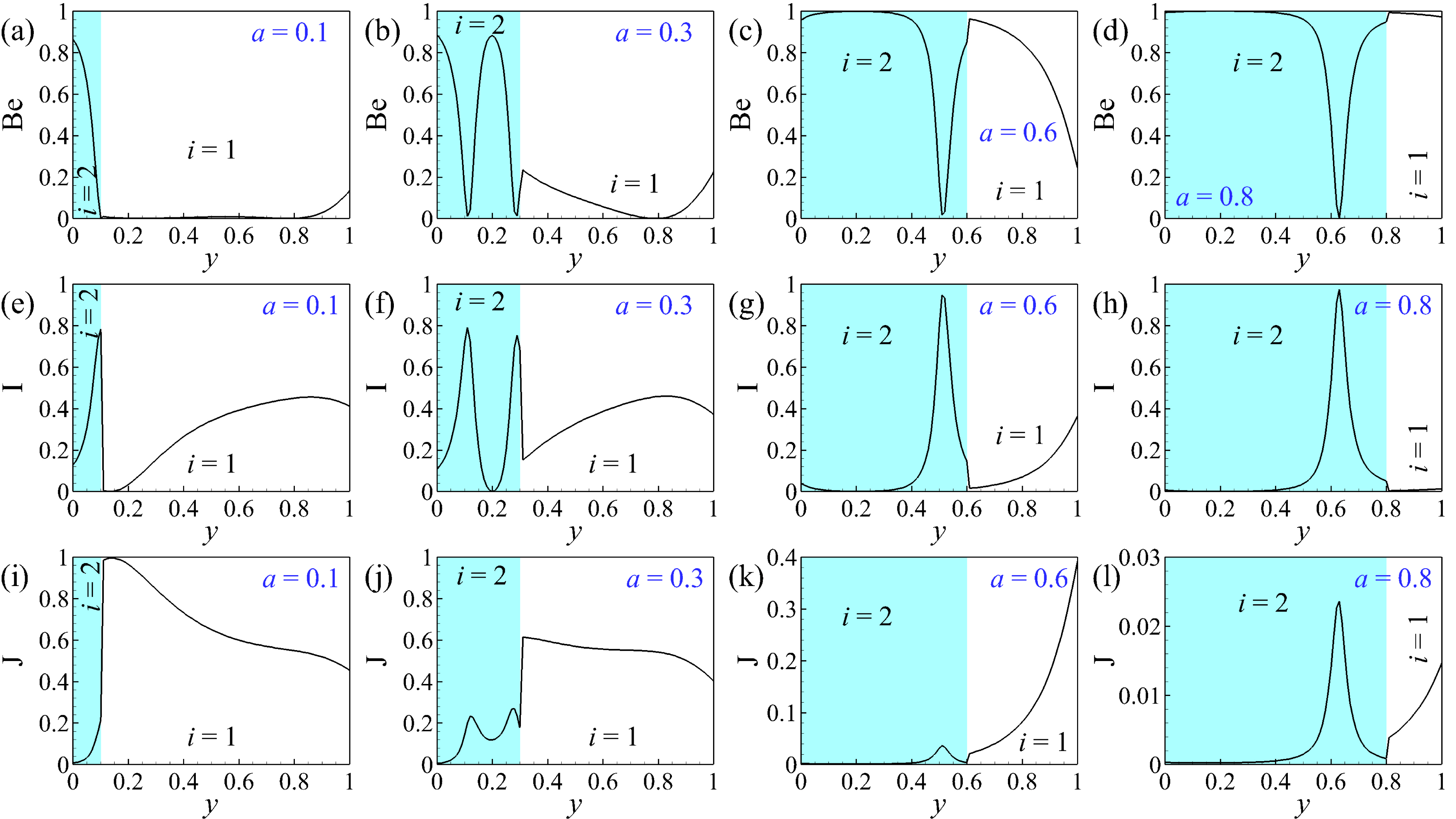}%
\caption{\label{fig3}Plots (a) $–$ (d), (e) $–$ (h), and (i) $–$ (l) show the variation of Bejan number (Be), magnetic field irreversibility parameter (I) and fluid flow irreversibility parameter (J) across the width (\textit{y}) of the channel for different filling ratios (\textit{a}), respectively. The other parameters are, \(\eta=2.0\), \(\mathrm{Ha}_1=3.0\), \(k=1.5\), \(\sigma=1.0\), \(\mathrm{Br}_1=2.0\), and \(F_1=2.0\).}
\end{figure*}
\indent The expressions for \(u_i\) and \(\theta_i\) [equations (\ref{25}) $–$ (\ref{26}) and (\ref{28}) $–$ (\ref{29})] show that, the velocity (\(u_i\)) of the fluids inside the channel essentially depends on \(a\), \(\eta\), and Ha\textsubscript{1} only, whereas, the temperature (\(\theta\)) distribution depends on the combined effect of all the seven dimensionless parameters as well as the velocity (\(u\)) profile inside the channel. For low filling ratios (\(a\) = 0.1 $–$ 0.6), the temperature distribution essentially depends on the velocity (\(u\)) profile of the fluids inside rather than the combined effect of the other dimensionless parameters such as, \(k\), \(\sigma\), Br\textsubscript{1}, and \(F_1\). However, for higher filling ratios, such as, in case of \(a\) = 0.8 (figure 2h), the effect of these dimensionless parameters (\(k\), \(\sigma\), Br\textsubscript{1}, and \(F_1\)) dictate the temperature distribution inside the channel rather than the velocity profile of the fluids alone. This is the reason behind the apparent decrease in maximum as well as average temperature inside the channel with the filling ratio, \(a\) = 0.8 in figure 2(h). \\
\indent Plots (a) $–$ (d), (e) $–$ (h), and (i) $–$ (l) in figure 3 show the variation of Bejan number (Be), magnetic field irreversibility parameter (I) and fluid flow irreversibility parameter (J) across the width (\textit{y}) of the channel for different filling ratios (\textit{a}), respectively. Plots (a) $–$ (d) show that with increase in filling ratio (\textit{a}), Be across the channel width (\textit{y}) increases. Plot (a) shows that, for \textit{a} = 0.1, the contribution in the entropy generation due to HTI or the Bejan number (Be) throughout the midsection of the channel (\(0.1\leq{y}\leq0.9\)) is negligible. With the increase in \textit{a}, Be across the channel width (\textit{y}) increases, however, there exists some minima across the width where the contribution in the entropy generation due to HTI are zero, such as, for (b) \textit{y} $\sim$ 0.1, $\sim$ 0.3, and $\sim$ 0.8, for (c) \textit{y} $\sim$ 0.5, and for (d) \textit{y} $\sim$ 0.6.\\
\indent Plots (e) $–$ (h) show that with increase in filling ratio (\textit{a}), I or the entropy generation due to MFI across the channel width (\textit{y}) decreases as an average, however, I\textsuperscript{\textit{max}} ($\sim$ 0.8 $–$ 1.0) remains similar. The locations of these maximum entropy generation due to MFI (I\textsuperscript{\textit{max}}) are same as the locations of the minima in case of HTI in plots (a) $–$ (d), which signifies that the across the channel width (\textit{y}) there are some zones where the entropy generation due to HTI (MFI) are highest (lowest) and vice versa. \\
\indent Plots (i) $–$ (l) show that, the entropy generation due to the FFI (or the dimensionless number J) decreases with the increase in filling ratio (\(a\)) across the width (\textit{y}) of the channel because of the less fluid flow as shown in figures 2(a) $–$ (d). However, it is interesting to note that, when the filling ratio (\textit{a}) is comparatively lower (= 0.1) in plot (i), the contribution in maximum entropy generation due to FFI ($\sim$ 1.0) near the interface is entirely due to the relatively unhindered fluid flow in the channel. Although with the increase in \textit{a}, the contribution to entropy generation due to the FFI (or J) near the fluid-fluid interface lowers [$\sim$ 0.6 for \textit{a} = 0.3 in plot (j) and $\sim$ 0.0 for \textit{a} = 0.6 and 0.8 in plots (k) and (l)] and is predominantly determined by the contribution from HTI (or Be). 
\subsection{Effect of viscosity ratio (\(\eta\))}
\begin{figure}
\includegraphics[width=1.0\linewidth]{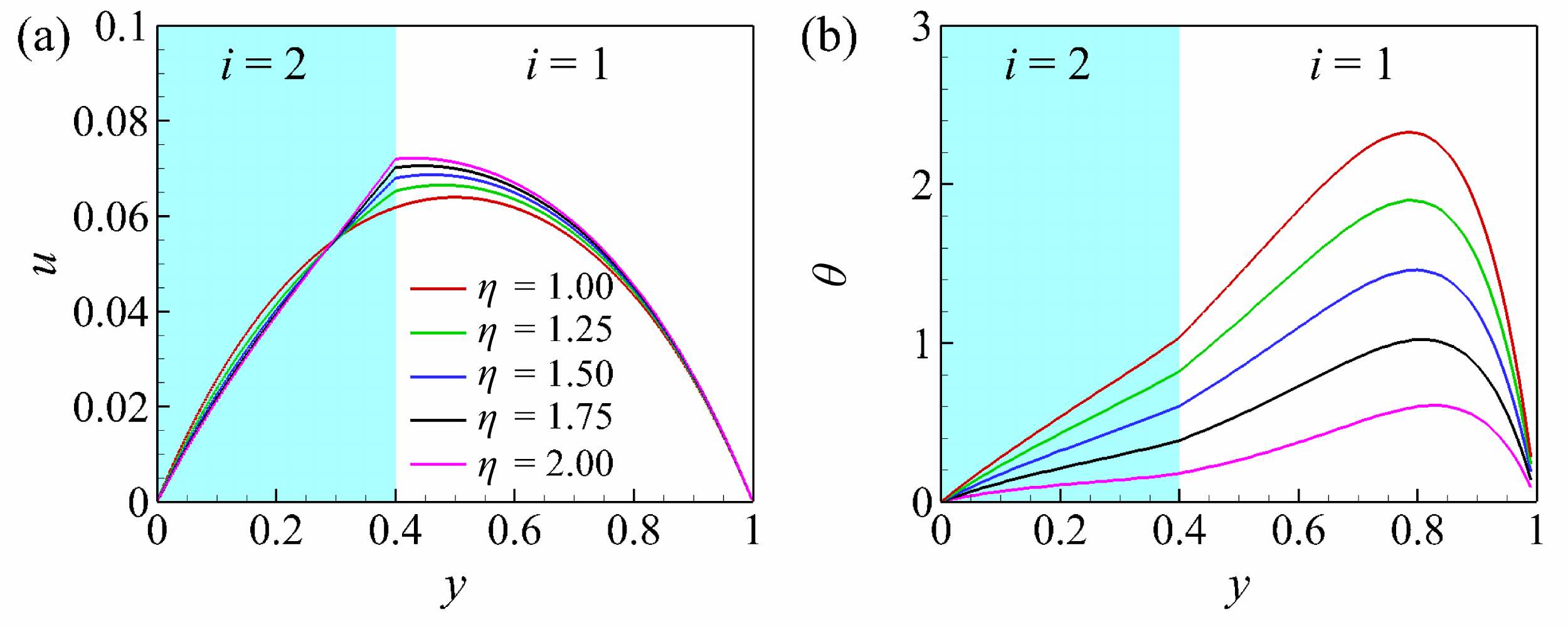}%
\caption{\label{fig4}Plots (a) and (b) show the variation of velocity (\(u\)) and temperature (\(\theta\)) across the width (\(y\)) of the channel for different viscosity ratios (\(\eta\)), respectively. The other parameters are, \(a=0.4\), \(\mathrm{Ha}_1=3.0\), \(k=1.5\), \(\sigma=1.0\), \(\mathrm{Br}_1=5.0\), and \(F_1=2.0\).}
\end{figure}
\indent Figures 4(a) and (b) display the variation of velocity (\(u\)) and temperature (\(\theta\)) along the width (\(y\)) of the channel with increasing viscosity ratios (\(\eta\)), respectively. First of all, when \(\eta\) = 1.0, the red line in plot (a) shows that the velocity profile is parabolic and consistent with the plane Poiseuille flow. Plot (a) also shows that, initially up to \(y\) $\sim$ 0.3, with the increase in \(\eta\), the velocity of fluid 2 is decreasing. However, beyond that the velocity of the fluids increase with increase in \(\eta\). Increasing the ratio \(\eta\) also decreases the dimensionless number Ha\textsubscript{2} (as, \(\displaystyle \mathrm{Ha}_2\propto1/\sqrt{\eta}\)), which augments the flow of fluid 2 inside the channel. In order to maintain the flow rate inside the channel due to the mass balance arising from the continuity equations [equations (5) and (9)], the velocity of fluid 1 increases with the increase in \(\eta\). At the fluid-fluid interface, the high velocity of the fluid 1 essentially increases the velocity of fluid 2 due to the laminar nature of the flow. \\
\indent With the increase in viscosity ratio (\(\eta\)), the temperature of both the fluids reduce as shown in figure 4(b). We have already discussed in figure 2 that, for low filling ratios (\(a\) = 0.1 $–$ 0.6), the temperature distribution essentially depends on the velocity (\(u\)) profile of the fluids inside rather than the combined effect of the other heat transfer related dimensionless parameters. In case of figure 4(b), since \(a\) = 0.4, the temperature profile of the fluids mainly depends on the velocity profile.\\

\indent Plot 4(a) shows that the velocity of fluid 1 is increasing with the increase in \(\eta\), whereas, the velocity of fluid 2 is slightly decreasing, which is why the temperature of the fluid 1 and 2 decreases with the increase in \(\eta\). It is noteworthy that, the velocity profile in plot (a) shows that the maximum velocity inside the channel is near the midline (\(y\) = 0.4 $–$ 0.6), whereas, the temperature profile shows that, the maximum temperature is near \(y\) = 0.8, which is near the upper wall of the channel. The maximum temperature inside the channel is dictated by the heat transfer mechanisms which are directly correlated to the combined effect of the dimensionless parameters, \(k\), \(\sigma\), Br\textsubscript{1}, and \(F_1\). \\
\begin{figure*}
\includegraphics[width=0.85\linewidth]{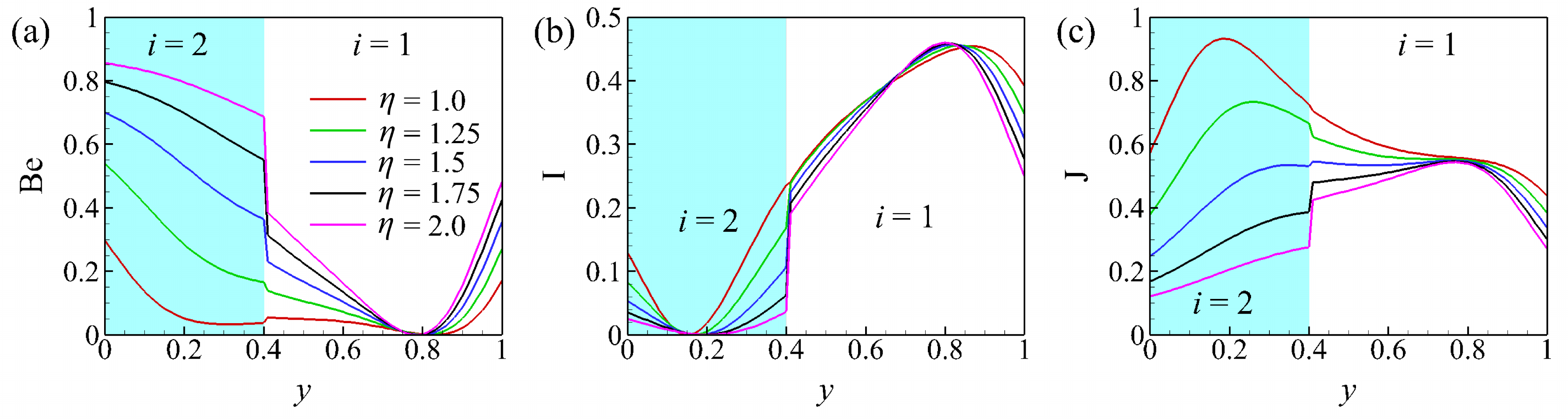}%
\caption{\label{fig5}Plots (a), (b), and (c) show the variation of Bejan number (Be), magnetic field irreversibility parameter (I) and fluid flow irreversibility parameter (J) across the width (\(y\)) of the channel for different viscosity ratios (\(\eta\)), respectively. The other parameters are, \(a=0.4\), \(\mathrm{Ha}_1=3.0\), \(k=1.5\), \(\sigma=1.0\), \(\mathrm{Br}_1=5.0\), and \(F_1=2.0\).}
\end{figure*}
\indent Figure 5(a) shows that, with the increase \(\eta\), the contribution in the entropy generation due to HTI or the Bejan number (Be) increases across the channel width (\(y\)). Since, increase in \(\eta\) from 1 to 2 reduces the flow of fluid 2 inside the channel, the entropy generation due to HTI in fluid 2 quadrupoles as shown in plot (a). However, the minima in plot (a) near \(y\) = 0.8 shows that the entropy generation due to HTI (or Be) at that location is negligible. It is noteworthy that, this is the same location where the temperature inside the channel is maximum as shown in figure 4(b). \\
\indent Figure 5(b) shows that, for a particular \(\eta\), the entropy generation due to MFI (or I) initially decreases, goes through a minimum near \(y = 0.2\), then increases, and goes through a maximum near \(y = 0.8\). It is also interesting to note that, I is always less than 0.5 throughout the channel width (\(y\)) for all the viscosity ratios (\(\eta\)), which signifies that, \(\mathrm{Be}+\mathrm{J}>\mathrm{I}\), or \(\mathrm{HTI}+\mathrm{FFI}>\mathrm{MFI}\). This also represents that, changing the viscosity ratio essentially changes the velocity and the temperature profile inside the fluids which in turn dictates the HTI and the FFI. However, the entropy generation due to the magnetic field (MFI) is not directly dependent on \(\eta\), rather it depends mainly on the flow and temperature profile change inside the channel due to the change in \(\eta\). \\
\indent Figure 5(c) depicts that, J decreases with the increase in \(\eta\) across the channel width (\(y\)). It can be seen from plot (c) that, when \(\eta\) is lower, the maximum contribution to entropy generation inside fluid 2 is due to the FFI compared to HTI and MFI. This is because the lesser value of \(\eta\) enhances the fluid flow inside the channel which in turn increases the entropy generation due to fluid flow. FFI goes through a maximum near \(y = 0.2\) in plot (c) for \(\eta = 1\) or 1.25. Plot (c) also shows a second local maxima near \(y = 0.8\) (fluid 1), where J $\sim$ 0.5 or more, which is the same location where, I also go through a local maximum as shown in plot (b). This signifies that, in the fluid 1 region near \(y = 0.8\), the entropy generation is mainly dictated by the combined effect from magnetic field (MFI) and the fluid flow (FFI) rather than the heat transfer component (HTI). \\
\indent Figure 5 shows that, the entropy generation inside the channel can be varied by altering the ratio \(\eta\) and the mechanism of entropy generation for both the fluids differ accordingly. The entropy generation due to FFI plays a major role for both the fluids (contribution \(\geq50\%\)), however, MFI (HTI) contributes less in case of fluid 2 (fluid 1) in the overall entropy generation inside the channel with the increase in viscosity ratio.
\subsection{Effect of thermal conductivity ratio (\(k\))}
\begin{figure}[h]
\includegraphics[width=0.7\linewidth]{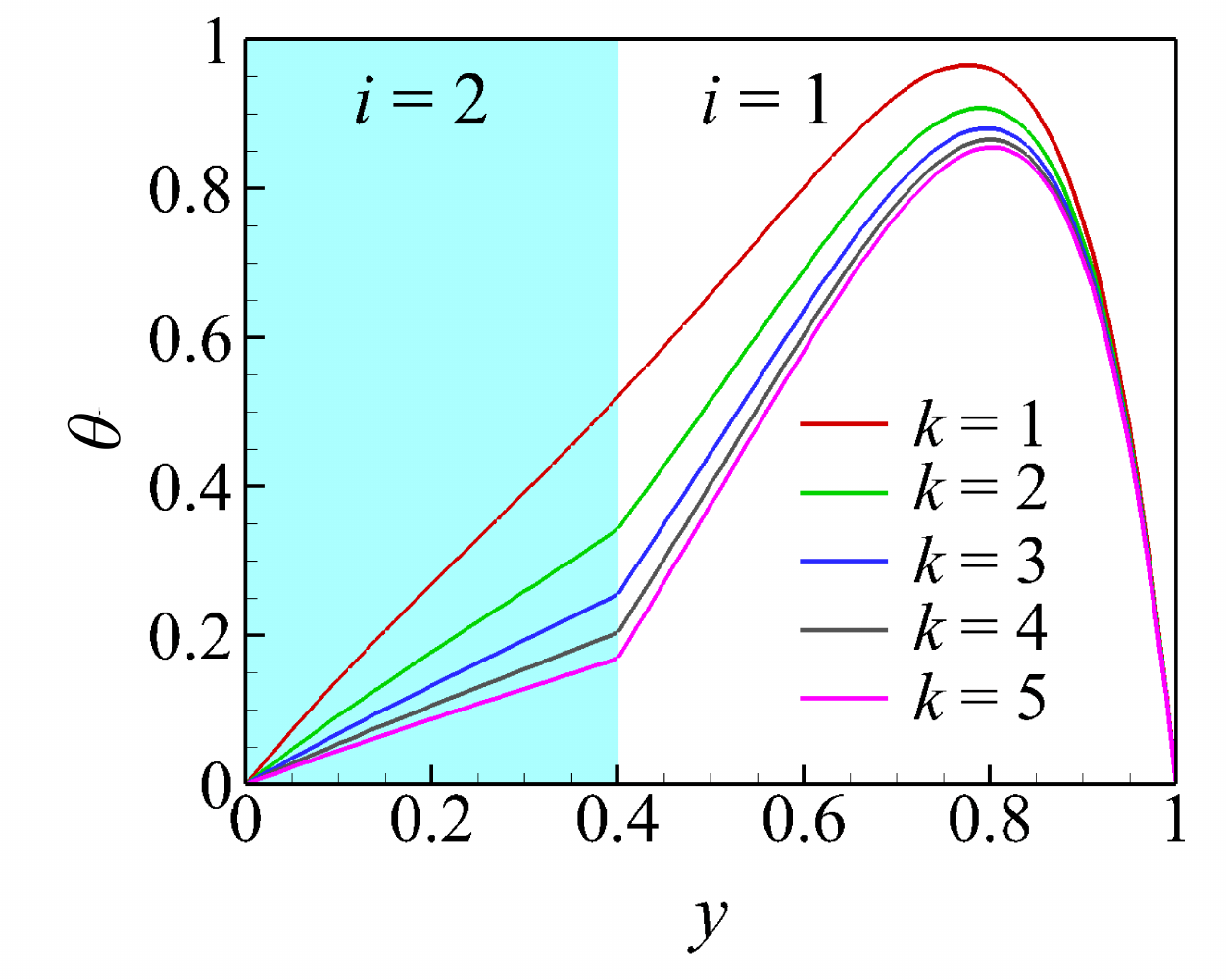}%
\caption{\label{fig6}Plot shows the temperature (\(\theta\)) distribution across the width (\(y\)) of the channel for different thermal conductivity ratios (\(k\)), respectively. The other parameters are, \(a=0.4\), \(\mathrm{Ha}_1=3.0\), \(\eta=2.0\), \(\sigma=1.0\), \(\mathrm{Br}_1=2.0\), and \(F_1=2.0\).}
\end{figure}
\begin{figure*}
\includegraphics[width=0.85\linewidth]{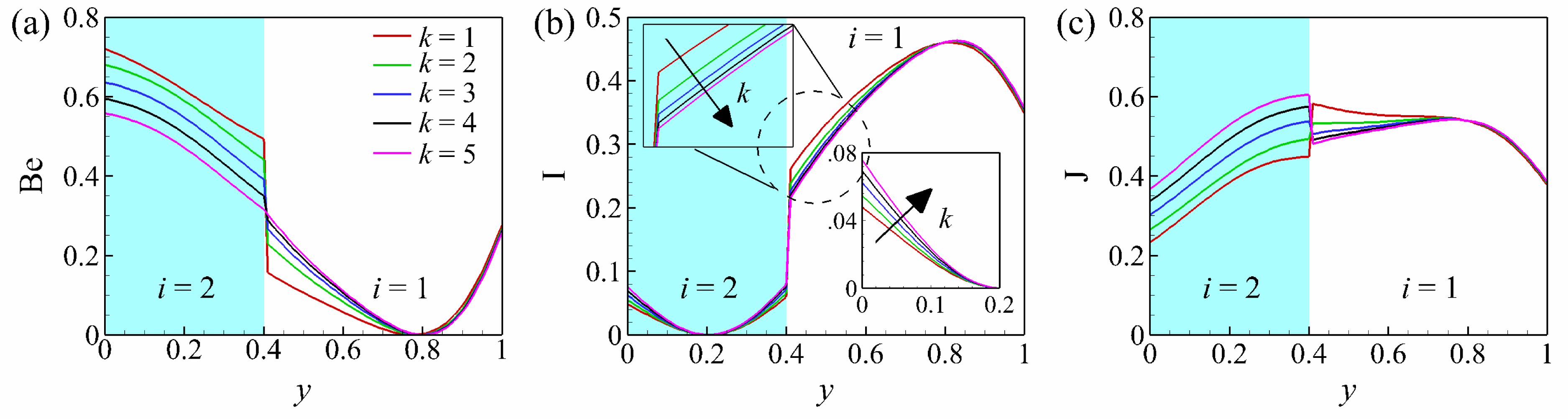}%
\caption{\label{fig7}Plots (a), (b), and (c) show the variation of Bejan number (Be), magnetic field irreversibility parameter (I) and fluid flow irreversibility parameter (J) across the width (\(y\)) of the channel for different thermal conductivity ratios (\(k\)), respectively. The other parameters are, \(a=0.4\), \(\mathrm{Ha}_1=3.0\), \(\eta=2.0\), \(\sigma=1.0\), \(\mathrm{Br}_1=2.0\), and \(F_1=2.0\).}
\end{figure*}
\indent Figure 6 shows the temperature profile across the width (\(y\)) of the channel for different thermal conductivity ratios (\(k\)). The heat transfer due to conduction inside the channel increases due to the increase in \(k\), which decreases the temperature (\(\theta\)) of the fluids as shown in figure 6. The thermal conductivity ratio (\(k\)) in our case is evaluated as the ratio of thermal conductivity of fluid 2 with fluid 1 (\(\displaystyle k=k_2/k_1\)). This ensures that at higher \(k\), the temperature of the fluid 2 (cyan region) decreases more rapidly than in fluid 1 (white region). Also, the maximum temperature of fluid 1 is more than fluid 2 because of the fluid flow conditions maintained by the flow parameters, \(\eta=2.0\), and \(\mathrm{Ha}_1=3.0\). The maximum temperature of the fluid 1 actually is near \(y = 0.8\), which is near the upper wall of the channel and similar to plot (b) in figure 4 due to the similar flow and thermal parameters.\\
\indent Figure 7(a) shows that, with the increase \(k\), the contribution in the entropy generation due to HTI or the Bejan number (Be) decreases across the channel width (\(y\)). Since, increase in \(k\) reduces the temperature distribution inside the channel, the entropy generation due to HTI in fluid 2 lowers significantly as shown in plot (a). Although, the value of Be for fluid 1 (Be\textsubscript{1}) always remains lower (\(\leq0.3\)) than that of the fluid 2 (Be\textsubscript{2} \(\geq0.3\)), but there is a stark contrast between the variation of Be\textsubscript{1} and Be\textsubscript{2}. It can be interpreted from figure 7(a) that, other than the global minima near \(y = 0.8\), Be\textsubscript{1} always increase with the increase in \(k\), although Be\textsubscript{2} decreases with the increase in k. However, the minima in plot (a) near \(y = 0.8\) shows that the entropy generation due to HTI (or Be) at that location is almost negligible. It is interesting to note that, this is the same location where the temperature inside the channel is maximum as shown in figure 6. \\
\indent Figure 7(b) shows that, for a particular \(k\), the entropy generation due to MFI (or I) initially decreases, goes through a minimum near \(y = 0.2\), then increases, and goes through a maximum near \(y = 0.8\). It is important to note that, I is always less than 0.5 throughout the channel width for all the thermal conductivity ratios (\(k\)), which signifies that, \(\mathrm{Be}+\mathrm{J}>\mathrm{I}\), or \(\mathrm{HTI}+\mathrm{FFI}>\mathrm{MFI}\). This also represents that, changing \(k\) essentially changes the temperature profile inside the fluids which in turn dictates the HTI and the FFI. However, the entropy generation due to the magnetic field (MFI) is not directly dependent on \(k\), rather it depends mainly on the flow and temperature profile change inside the channel due to the change in \(k\). Furthermore, the entropy generation due to MFI (or I) near the fluid-fluid interface suddenly increases greater than \(\sim 10\%\) from fluid 2 (\(I_2<0.1\)) to fluid 1 (\(I_1>0.2\)) irrespective of the value to \(k\), which is also an important factor to consider the irreversibility due to the presence of a fluid-fluid interface in figure 7(b). Figure 7(b) also shows that, other than the global minima (maxima) near \(y = 0.2 (0.8)\), \(I_1\)(\(I_2\)) always decrease (increase) with the increase in \(k\), as shown clearly in the inset of figure 7(b).\\
\indent Figure 7(c) depicts that \(J_1\)(\(J_2\)) always decrease (increase) with the increase in \(k\), across the channel width. It can be seen from plot (c) that, when \(k\) is lower (1 for example), the maximum contribution to entropy generation inside fluid 2 is due to HTI (contribution $\sim$ 70\%) compared to FFI (contribution $\sim$ 20\%) and MFI (contribution $\sim$ 10\%). This is because the lesser value of \(k\) enhances the temperatures of the fluids inside the channel which in turn increases the entropy generation due to HTI.  Plot (c) also shows that irrespective of the fluids, the contribution to the total entropy generation inside the channel due to FFI is always within 20\% $-$ 60\%. This signifies that even though the thermal conductivity ratio does not necessarily change the fluid flow inside the channel as such, however, a significant portion of the total entropy generation due to fluid flow can be controlled by controlling \(k\). \\
\indent In summary, figure 7 shows that, due to the change in \(k\), the entropy generation inside the channel can be varied, also, the mechanism of entropy generation for both the fluids differ significantly. The entropy generation due to HTI plays a major role for fluid 2 (contribution \(\geq40\%\)), however, in case of fluid 1 the overall entropy generation is dictated by FFI (contribution \(\geq40\%\)) mainly.
\subsection{Effect of electrical conductivity ratio (\(\sigma\))}
\begin{figure}[h]
\includegraphics[width=0.7\linewidth]{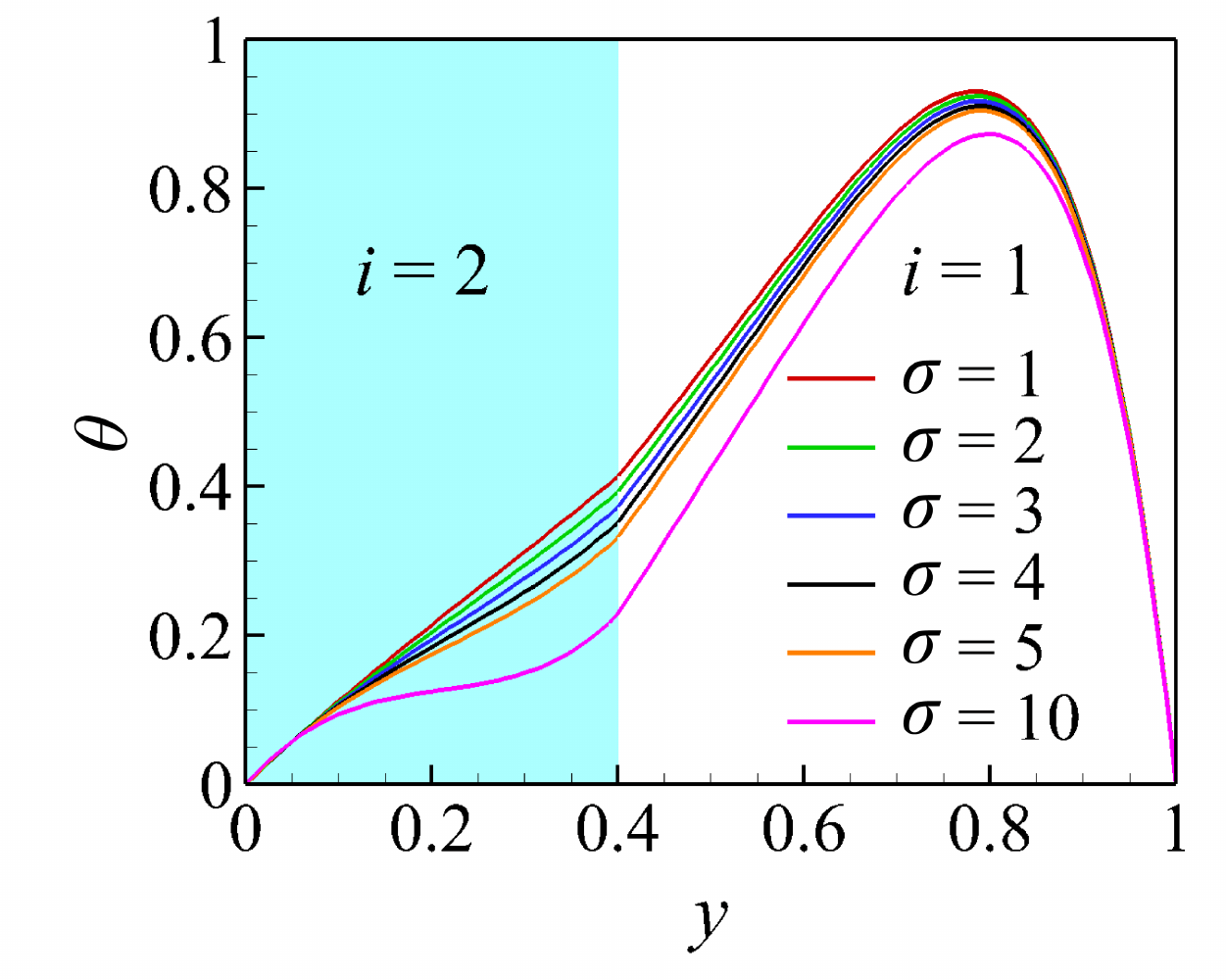}%
\caption{\label{fig8}Plot shows the temperature (\(\theta\)) distribution across the width (\(y\)) of the channel for different electrical conductivity ratios (\(\sigma\)), respectively. The other parameters are, \(a=0.4\), \(\mathrm{Ha}_1=3.0\), \(\eta=2.0\), \(k=1.5\), \(\mathrm{Br}_1=2.0\), and \(F_1=2.0\).}
\end{figure}
\indent Figure 8 shows the temperature profile across the width (\(y\)) of the channel for different electrical conductivity ratios (\(\sigma\)). Increasing the ratio \(\sigma\) increases the dimensionless number \(\mathrm{Ha}_2\) (as, \(\displaystyle \mathrm{Ha}_i\propto{\sqrt{\sigma_i}}\)), which hinders the flow of fluid 2 inside the channel. When the velocity of fluid 2 decreases inside the channel the heat transfer via radiative mode increases. The plot in figure 8 shows that, this low fluid velocity facilitates the considerable temperature drop (from \(\displaystyle \theta_2^{avg} \sim 0.3\) to \(\displaystyle \theta_2^{avg} \sim 0.1\), \(\displaystyle \Delta\theta_2^{avg} \sim 0.2\)) in fluid 2, whereas, the drop in fluid 1 temperature is considerably lower (\(\displaystyle \Delta\theta_2^{avg} \leq 0.1\)). \\
\begin{figure*}
\includegraphics[width=0.85\linewidth]{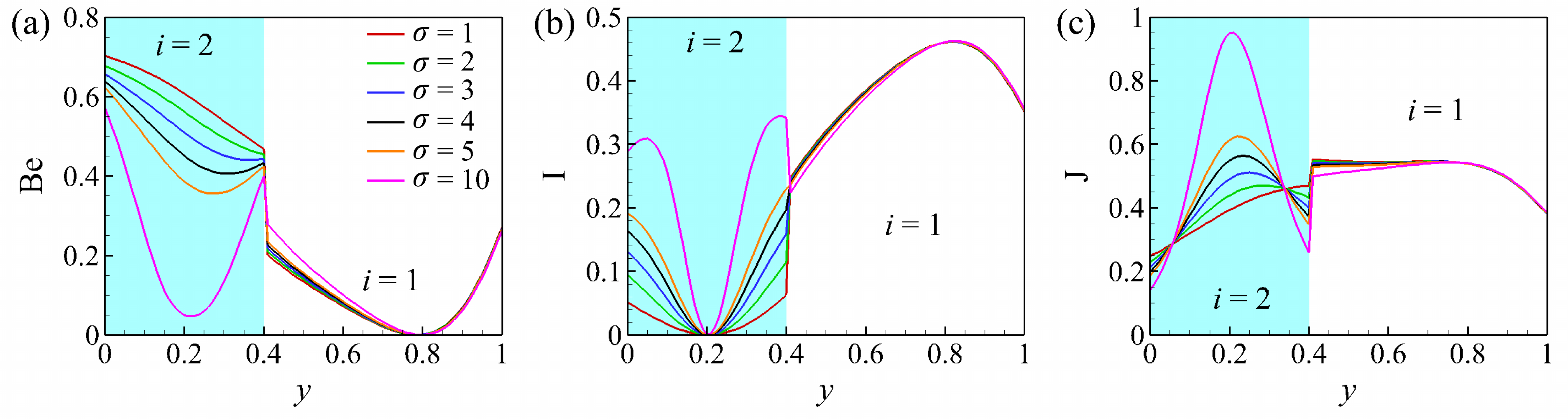}%
\caption{\label{fig9}Plots (a), (b), and (c) show the variation of Bejan number (Be), magnetic field irreversibility parameter (I) and fluid flow irreversibility parameter (J) across the width (\(y\)) of the channel for different electrical conductivity ratios (\(\sigma\)), respectively. The other parameters are, \(a=0.4\), \(\mathrm{Ha}_1=3.0\), \(\eta=2.0\), \(k=1.5\), \(\mathrm{Br}_1=2.0\), and \(F_1=2.0\).}
\end{figure*}
\indent Figure 9(a) shows that, with the increase in \(\sigma\), the contribution in the entropy generation due to HTI or the Bejan number (Be) decreases in fluid 2. Since, increase in \(\sigma\) reduces the temperature distribution inside the channel, the entropy generation due to HTI in fluid 2 lowers significantly as shown in plot (a). Although, the value of Be for fluid 1 (\(\mathrm{Be}_1\)) always remains lower (\(\leq0.25\)) than that of the fluid 2 (\(\mathrm{Be}_2\geq0.35\)) for most of the cases (\(\sigma\leq5.0\)), but there is a significant contrast between the variation of \(\mathrm{Be}_1\) and \(\mathrm{Be}_2\) with the variation in \(\sigma\). It can be interpreted from figure 9(a) that, other than the global minima near \(y = 0.8\), the variation of \(\mathrm{Be}_1\) always remains similar with the increase in \(\sigma\), although \(\mathrm{Be}_2\) decreases with the increase in \(\sigma\). The global minima in plot (a) near \(y = 0.8\) shows that the entropy generation due to HTI (or Be) at that location is almost negligible. It is important to note that, this is the same location where the temperature inside the channel is maximum as shown in figure 8.\\
\indent Another interesting observation from figure 9(a) is that, in fluid 2, \(\mathrm{Be}_2\) initially decreases with the increase in \(y\), and eventually increases with further increase in \(y\). Also, with the increase in \(\sigma\), the rate of decrease or increase (minimum \(\mathrm{Be}_2\), \(\mathrm{Be}_2^{min}\)) of \(\mathrm{Be}_2\) in fluid 2 increases (decreases) as shown in plot (a). For example, \(\mathrm{Be}_2^{min}\sim0.4\) at \(y = 0.3\) for \(\sigma=4.0\) in fluid 2 gets shifted to \(\mathrm{Be}_2^{min}\sim0.05\) at \(y = 0.2\) for \(\sigma=10.0\). \\
\indent Figure 9(b) shows that, for all \(\sigma\) (= 1 $-$ 10), the entropy generation due to MFI in fluid 1 (or \(\mathrm{I}_1\)) initially increases, goes through a maximum near \(y = 0.8\), then decreases, although the magnitude of \(\mathrm{I}_1\) vary minimally across the width of the channel (\(a\leq{y}\leq1\)) for increasing \(\sigma\). However, increasing \(\sigma\) changes the distribution of \(\mathrm{I}_2\) across the width of the channel (\(0\leq{y}\leq{a}\)) for fluid 2. For almost all \(\sigma\) (= 1 $-$ 5), \(\mathrm{I}_2\) initially decreases with increase in \(y\), then goes through a minimum near \(y = 0.2\), and then increases with further increase in \(y\) till \(y\leq{a}\).\\
\indent Also, plot (b) shows that, near the wall (\(y = 0\)) and the interface of fluid 1 and 2 (\(y = a\)), the magnitude of \(\mathrm{I}_2\) remains maximum and this maximum magnitude increases with the increase in \(\sigma\). This distribution implies that, with the increase in electrical conductivity of fluid 2, the entropy generation due to MFI increases near the wall and the interface. However, with further increase in \(\sigma\left(=10\right)\) the variation \(\mathrm{I}_2\) changes significantly across fluid 2 (\(0\leq{y}\leq{a}\)). The variation of \(\mathrm{I}_2\) for \(\sigma=10\) in figure 9(b) shows that, near the lower wall (\(y = 0\)) and the fluid-fluid interface (\(y = a\)), the magnitude of \(\mathrm{I}_2\) briefly goes through a pair of local maximums at \(y = 0.05\) and \(y = 0.35\). This apparent shift of local maximums from the wall (\(y = 0\)) and the interface (\(y = a\)) towards the inner regions of fluid 2 (\(y = 0.05\) and \(y = 0.35\)) implies that, with the increase in orders of magnitude of \(\sigma\) (1 to 10) the entropy generation due to MFI shifts towards the bulk of the fluid 2. It is also interesting to note that, I is always less than 0.5 throughout the channel width for all \(\sigma\) (= 1 $-$ 10), which signifies that, Be+J $>$ I, or HTI+FFI $>$ MFI.  \\
\indent Figure 9(c) illustrates that \(\mathrm{J}_1\) always remains similar with the increase in \(\sigma\), although \(\mathrm{J}_2\) varies significantly with the increase in \(\sigma\). It can be seen from plot 9(c) that, for \(\sigma=1\),  \(\mathrm{J}_2\) increases monotonically with the increase in \(y\) up to \(y\leq{a}\), whereas, for \(\sigma=2-10\), \(\mathrm{J}_2\) initially increases monotonically with the increase in \(y\) up to a certain value and then decreases with further increase in \(y\) up to \(y\leq{a}\). This distribution ensures that, near the wall (\(y = 0\)) and the fluid-fluid interface (\(y = a\)) the entropy generation due to FFI (or \(\mathrm{J}_2\)) actually decreases with the increase in \(\sigma\).  Also, figure 9(c) shows that, the maximum \(\mathrm{J}_2\)(\(\mathrm{J}_2^{max}\)) shifts minutely from \(y>a/2\) towards \(y\sim{a/2}\) with the increase \(\sigma\) from 2 to 10.\\
\indent In short, figure 9 shows that, the magnitudes and mechanisms of entropy generation inside the channel can be altered by changing the \(\sigma\) for both the fluids . For most of the cases, the entropy generation due to FFI and HTI plays a major role for fluid 2 (total contribution \(\geq70\%\)), however, in case of fluid 1 the overall entropy generation is mainly dictated by FFI and MFI (total contribution \(\geq70\%\)).
\subsection{Effect of Hartmann number (Ha\textsubscript{1})}
\begin{figure}[h!]
\includegraphics[width=1.0\linewidth]{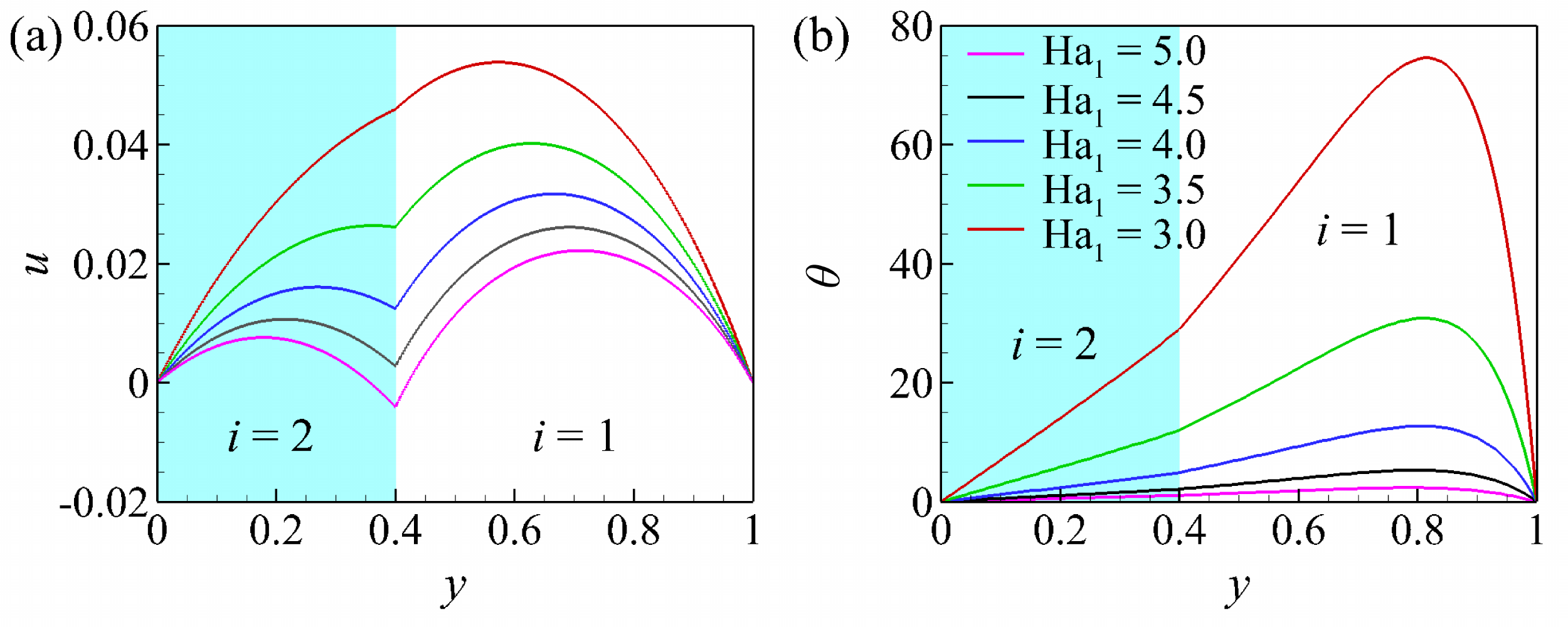}%
\caption{\label{fig10}Plots (a) and (b) show the variation of velocity (\(u\)) and temperature (\(\theta\)) distribution across the width (\(y\)) of the channel for different Hartmann numbers (\(\mathrm{Ha}_1\)), respectively. The other parameters are, \(a=0.4\), \(\eta=2.0\), \(k=1.5\), \(\sigma=1.0\), \(\mathrm{Br}_1=5.0\), and \(F_1=2.0\).}
\end{figure}
\begin{figure*}
\includegraphics[width=0.85\linewidth]{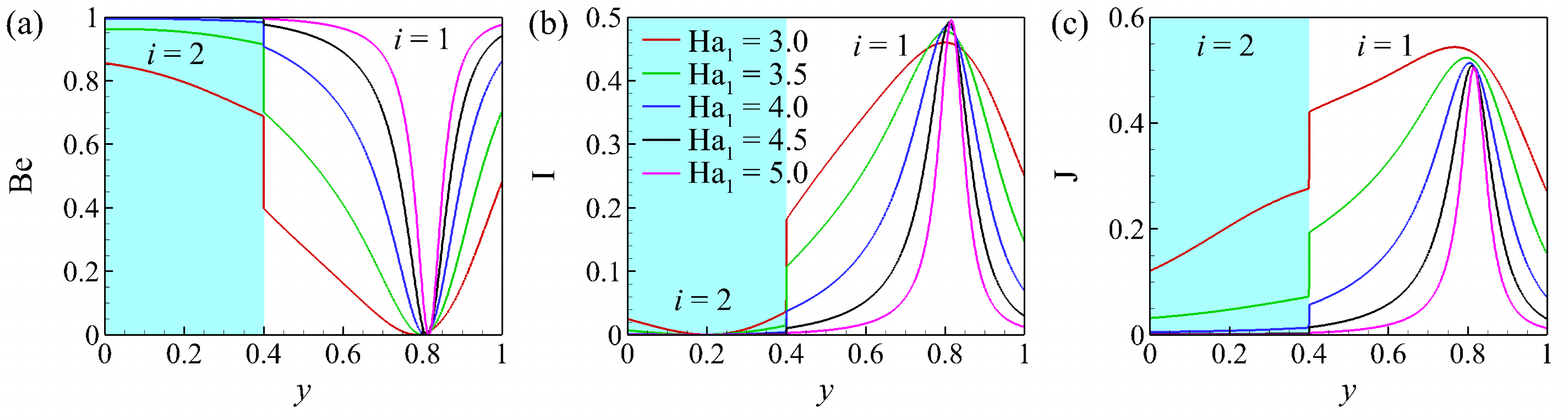}%
\caption{\label{fig11}Plots (a), (b), and (c) show the variation of Bejan number (Be), magnetic field irreversibility parameter (I) and fluid flow irreversibility parameter (J) across the width (\(y\)) of the channel for different Hartmann numbers (\(\mathrm{Ha}_1\)), respectively. The other parameters are, \(a=0.4\), \(\eta=2.0\), \(k=1.5\), \(\sigma=1.0\), \(\mathrm{Br}_1=5.0\), and \(F_1=2.0\).}
\end{figure*}
Figures 10(a) and 10(b) show the variation of velocity (\(u\)) and temperature (\(\theta\)) along the width (\(y\)) of the channel with increasing Hartmann numbers (\(\mathrm{Ha}_1\)), respectively. Increase in \(\mathrm{Ha}_1\), increases the effect of Lorentz force inside the channel, which acts as a resistance to the flow. Therefore, it is observed that the velocity in the channel decreases with the increase in the value of \(\mathrm{Ha}_1\). Duwairi \textit{et al.} showed in a previous study that for a single fluid under the influence of a transverse magnetic field, the magnitudes of velocity and temperature distribution decrease with the increase in Ha \cite{duwairi2007}. This means that the electrical conductivity of the fluid dictates the velocity and the temperature profiles. We have also modified the present mathematical formulation to asymptotically match with the governing equations of Duwairi \textit{et al.} by employing the same fluid properties for both the fluids in order to mimic a single fluid system \cite{duwairi2007}. The results from this asymptotic study shows a good match with the results of Duwairi \textit{et al.} \cite{duwairi2007}. The comparison of these results are shown in figure \ref{fig17} of appendix B. Even in bilayer flows, the similar trend of reducing velocity and temperature upon increasing Ha is observed in the present study as shown in figure \ref{fig10}. \\
\indent Another important thing to note is that, for \(\mathrm{Ha}_1=5.0\), the velocity of the fluids at the interface becomes \(u_i\leq0\), this signifies that for \(\mathrm{Ha}_1>5.0\) the velocity of the fluids near the interfacial region can theoretically become negative. The negative velocity of the fluids near the interfacial region \(\mathrm{Ha}_1>5.0\) signifies that the fluids in that region may backflow inside the channel upon further increase in \(\mathrm{Ha}_1\). This back-flow near the fluid-fluid interface may induce circulatory motion of the fluids near the interfacial region. With the increase in \(\mathrm{Ha}_1\), the temperature of the fluids inside the channel decreases as shown in figure 10(b). The temperature of the fluids declines because of the increase in radiative heat transfer due to the slow-moving fluids inside the channel as a result of increased \(\mathrm{Ha}_1\).\\
\indent Figure 11(a) shows that, in general, Be increases across the width of the channel with the increase in \(\mathrm{Ha}_1\). For fluid 2, increase in \(\mathrm{Ha}_1\) from 3 to 5 increases \(\mathrm{Be}_2\) up to 1, which signify that, the contribution of HTI to the total entropy generation in fluid 2 is essentially 100\%. Although, increase in \(\mathrm{Ha}_1\) from 3 to 5 increases \(\mathrm{Be}_1\) too however, the minima in plot (a) near \(y = 0.8\) shows that the entropy generation due to HTI (or \(\mathrm{Be}_1\)) at that location is negligible for fluid 1. This is because, the temperature and the velocity of fluids inside the channel is maximum at this same location as shown in figure 10. \\
\indent Figure 11(b) shows that, magnetic field irreversibility parameter (\(\mathrm{I}_2\)) in fluid 2 decreases and reaches almost zero with the increase in \(\mathrm{Ha}_1\). Across the width of the channel (\(0\leq{y}\leq{a}\)), \(\mathrm{I}_2\) initially decreases, goes through a minimum near \(y = 0.2\), and then eventually increases near the fluid-fluid interface near \(y = a\). However, the magnitude of \(\mathrm{I}_2\) is so insignificant (\(\leq0.02\)) in fluid 2 that its contribution to entropy generation can be ignored. In case of fluid 1, \(\mathrm{I}_1\) also decreases with the increase in \(\mathrm{Ha}_1\), however, for a particular \(\mathrm{Ha}_1\) and across the width of the channel (\(a\leq{y}\leq{1}\)) it initially increases, goes through a maxima near \(y = 0.8\), and then eventually decreases near the channel wall. \\
\indent Interestingly, the peak of \(\mathrm{I}_1\) in plot (b) shows that the contribution to the entropy generation due to MFI near \(y = 0.8\), can be as high as $\sim$ 50\%, whereas, plot (a) shows that the contribution to the entropy generation due to HTI near \(y = 0.8\), can be as low as $\sim$ 0\%. This is because, increase in \(\mathrm{Ha}_1\), increases the effect of Lorentz force inside the channel, which acts as a resistance to the flow. The increase in radiative heat transfer due to the slow-moving fluids inside the channel as a result of increased   \(\mathrm{Ha}_1\) ensures that the HTI plays a significantly major role in entropy generation inside the channel for the most part. The highest temperature of fluids near \(y = 0.8\) although guarantees that, at that location, the entropy generation is dictated by the other two irreversibilities (MFI and FFI). \\ 
\indent Figure 11(c) shows that, J decreases across the width of the channel with the increase in \(\mathrm{Ha}_1\), as the velocity of the fluids inside the channel decreases due to the hinderance from the increased Lorentz force. With the increase in \(\mathrm{Ha}_1\) from 3 to 5, average \(\mathrm{J}_2\left(\mathrm{J}_2^{avg}\right)\) decreases from $\sim$ 0.2 to zero. Interestingly, the peak of \(\mathrm{J}_1\left(\mathrm{J}_1^{max}\right)\) in plot (c) shows that the contribution to the entropy generation due to FFI near \(y = 0.8\), can be as high as $\sim$ 50\%. \\
\indent Moreover, figure 11 shows that, change in \(\mathrm{Ha}_1\) significantly alters the magnitude of the entropy generation inside the channel as well as the mechanism of entropy generation for both the fluids. Majority of entropy generation inside the region of fluid 2 is dictated by the HTI ($\sim$ 80\% $-$ 100\%) with the increase in \(\mathrm{Ha}_1\). However, the contribution to the total entropy generation due to the three irreversibilities are shared for fluid 1 for the most part, except near the upper wall (\(y = 0.8\)), where entropy generation due to HTI is negligible. At this location, the entropy generation is almost equally shared by both MFI (contribution  $\sim$ 50\%) and FFI (contribution $\sim$ 50\%). 
\subsection{Effect of Brinkman number (Br\textsubscript{1})}
\begin{figure}[h]
\includegraphics[width=0.7\linewidth]{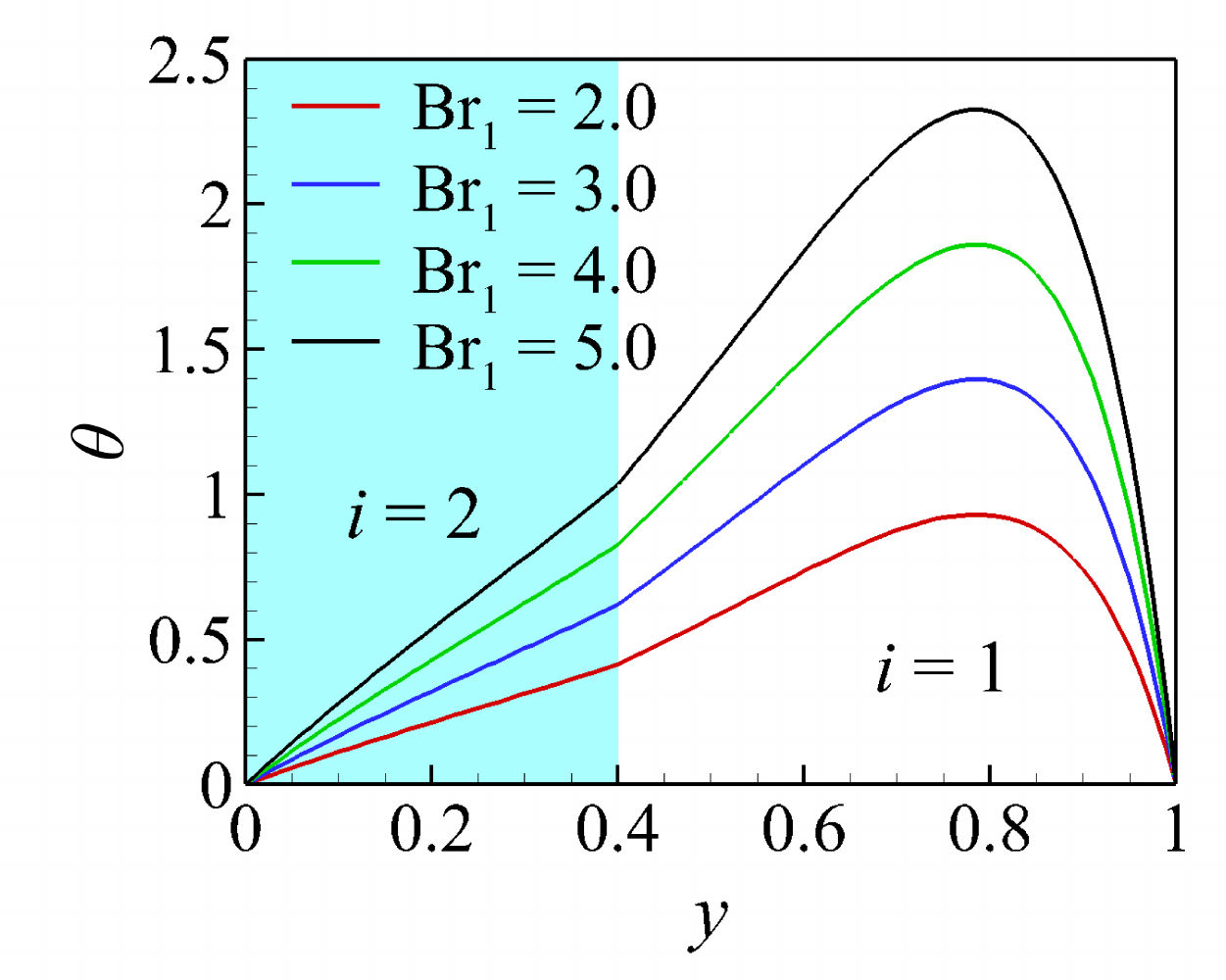}%
\caption{\label{fig12}Plot shows the temperature (\(\theta\)) distribution across the width (\(y\)) of the channel for different Brinkman numbers (\(\mathrm{Br}_1\)), respectively. The other parameters are, \(a=0.4\), \(\eta=2.0\), \(k=1.5\), \(\sigma=1.0\), \(\mathrm{Ha}_1=5.0\), and \(F_1=2.0\).}
\end{figure}
\begin{figure*}[t]
\includegraphics[width=0.85\linewidth]{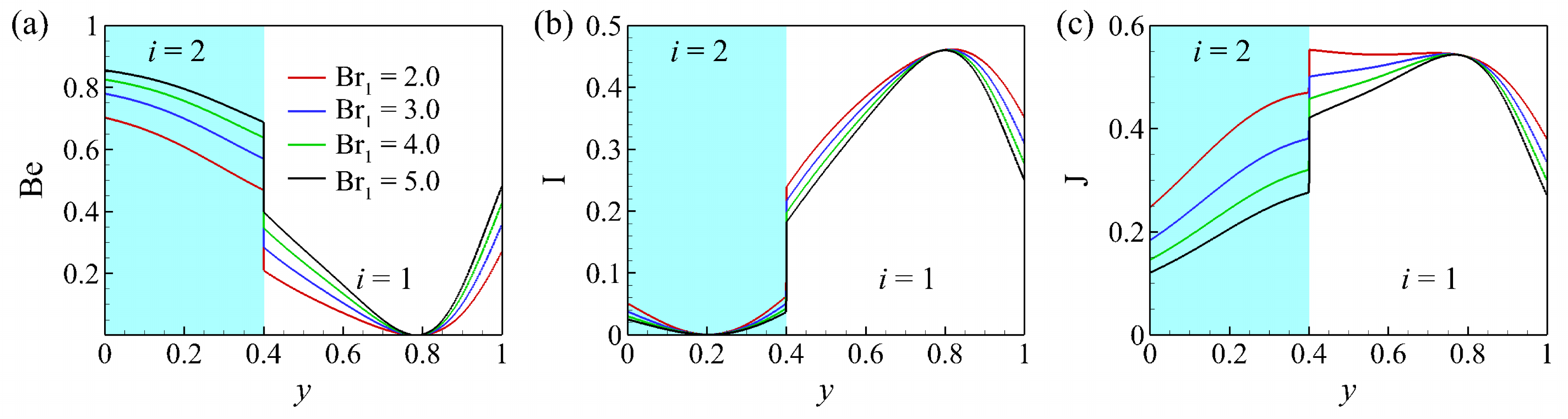}%
\caption{\label{fig13}Plots (a), (b), and (c) show the variation of Bejan number (Be), magnetic field irreversibility parameter (I) and fluid flow irreversibility parameter (J) across the width (\(y\)) of the channel for different Brinkman numbers (\(\mathrm{Br}_1\)), respectively. The other parameters are, \(a=0.4\), \(\eta=2.0\), \(k=1.5\), \(\sigma=1.0\), \(\mathrm{Ha}_1=5.0\), and \(F_1=2.0\).}
\end{figure*}
Figure 12 illustrates the temperature profile across the width (\(y\)) of the channel for increasing Brinkman numbers (\(\mathrm{Br}_1\)). Brinkman numbers (\(\mathrm{Br}_i\)) signify the ratio between heat produced by viscous dissipation and heat transported by molecular conduction. With the increase in \(\mathrm{Br}_1\), the heat produced inside the channel by viscous dissipation increases compared to the heat transferred due to conduction, which facilitates the increase in temperature or both the fluids inside the channel as shown in figure 12.\\
\indent Figure 13(a) shows that, with the increase in \(\mathrm{Br}_1\), the contribution in the entropy generation due to HTI or Bejan number (Be) increases across the channel width (\(y\)). Since, increase in \(\mathrm{Br}_1\) increases the temperature distribution inside the channel, the entropy generation due to HTI in fluid 2 significantly increases as shown in plot (a). Although, the value of Be for fluid 1 (\(\mathrm{Be}_1\)) always remains lower (\(0\leq{\mathrm{Be}_1}\leq0.4\)) than that of the fluid 2 (\(0.4\leq{\mathrm{Be}_2}\leq0.85\)), but there is a visible contrast between the variation of \(\mathrm{Be}_1\) and \(\mathrm{Be}_2\). Across the channel width (\(0\leq{y}\leq{a}\)), \(\mathrm{Be}_2\) always decreases with the increase in \(y\), whereas, \(\mathrm{Be}_1\) across the channel width (\(a\leq{y}\leq{1}\)), initially decreases, goes through a minima \(y = 0.8\), then eventually increases near the channel upper wall with the increase in \(y\). It can be interpreted from figure 13(a) that, other than the global minima near \(y = 0.8\), \(\mathrm{Be}_1\) always increase with the increase in \(\mathrm{Br}_1\). However, the minima in plot (a) near \(y = 0.8\) shows that the entropy generation due to HTI (or Be) at that location is almost negligible. It is interesting to note that, this is the same location where the temperature inside the channel is maximum as shown in figure 12. \\
\indent Figure 13(b) shows that, for a particular \(\mathrm{Br}_1\), the entropy generation due to MFI (or I) initially decreases, goes through a minimum near \(y = 0.2\), then increases, and goes through a maximum near \(y = 0.8\). It is also interesting to note that, I is always less than 0.5 throughout the channel width for all \(\mathrm{Br}_1\), which signifies that, Be+J \(>\) I, or HTI+FFI \(>\) MFI. This also represents that, varying \(\mathrm{Br}_1\) essentially changes the temperature profile inside the fluids which in turn dictates the HTI and the FFI. However, the entropy generation due to the magnetic field (MFI) is not directly dependent on \(\mathrm{Br}_1\), rather it depends mainly on the flow and temperature profile change inside the channel due to the change in \(\mathrm{Br}_1\). Also, the entropy generation due to MFI (or I) near the fluid-fluid interface suddenly increases greater than $\sim$ 15\% from fluid 2 (\(\mathrm{I}_2<0.05\)) to fluid 1 (\(\mathrm{I}_1>0.2\)) irrespective of the value of \(\mathrm{Br}_1\), which is also an important factor to consider in figure 13(b). Figure 13(b) also shows that, other than the global minima (maxima) near \(y\) = 0.2 (0.8), both \(\mathrm{I}_1\) and \(\mathrm{I}_2\) always decrease with the increase in \(\mathrm{Br}_1\), as shown clearly in figure 13(b). \\
\indent Figure 13(c) depicts that, both \(\mathrm{J}_1\) and \(\mathrm{J}_2\) always decrease with the increase in \(\mathrm{Br}_1\), across the channel width. It can be seen from plot (c) that, when \(\mathrm{Br}_1\) is lower (2 for example), the maximum contribution to entropy generation inside fluid 2 is due to HTI (contribution $\sim$ 70\%) compared to FFI (contribution $\sim$ 25\%) and MFI (contribution $\sim$ 5\%). This is because the lesser value of \(\mathrm{Br}_1\) enhances the molecular conduction inside the channel which in turn increases the entropy generation due to HTI. Plot (c) also shows that irrespective of the fluids the contribution to the total entropy generation inside the channel due to FFI is always within 15\% $-$ 60\%. This signifies that, even though \(\mathrm{Br}_1\) does not necessarily change the fluid flow inside the channel as such, however, a significant portion of the total entropy generation which is due to the FFI can be controlled by monitoring \(\mathrm{Br}_1\).\\
\indent In summary, figure 13 shows that, due to the change in \(\mathrm{Br}_1\), the entropy generation inside the channel can be tuned by altering the mechanism of entropy generation for both the fluids. The entropy generation due to HTI plays a major role for fluid 2 (contribution \(\geq50\%\)), however, in case of fluid 1 the overall entropy generation is dictated by all the three irreversibilities (HTI $\sim$ 0\% $–$ 40\%, MFI $\sim$ 20\% $-$ 45\%, and FFI $\sim$ 35\% $–$ 55\%).
\subsection{Effect of radiation parameter (\(F_1\))}
\begin{figure}[h!]
\includegraphics[width=0.7\linewidth]{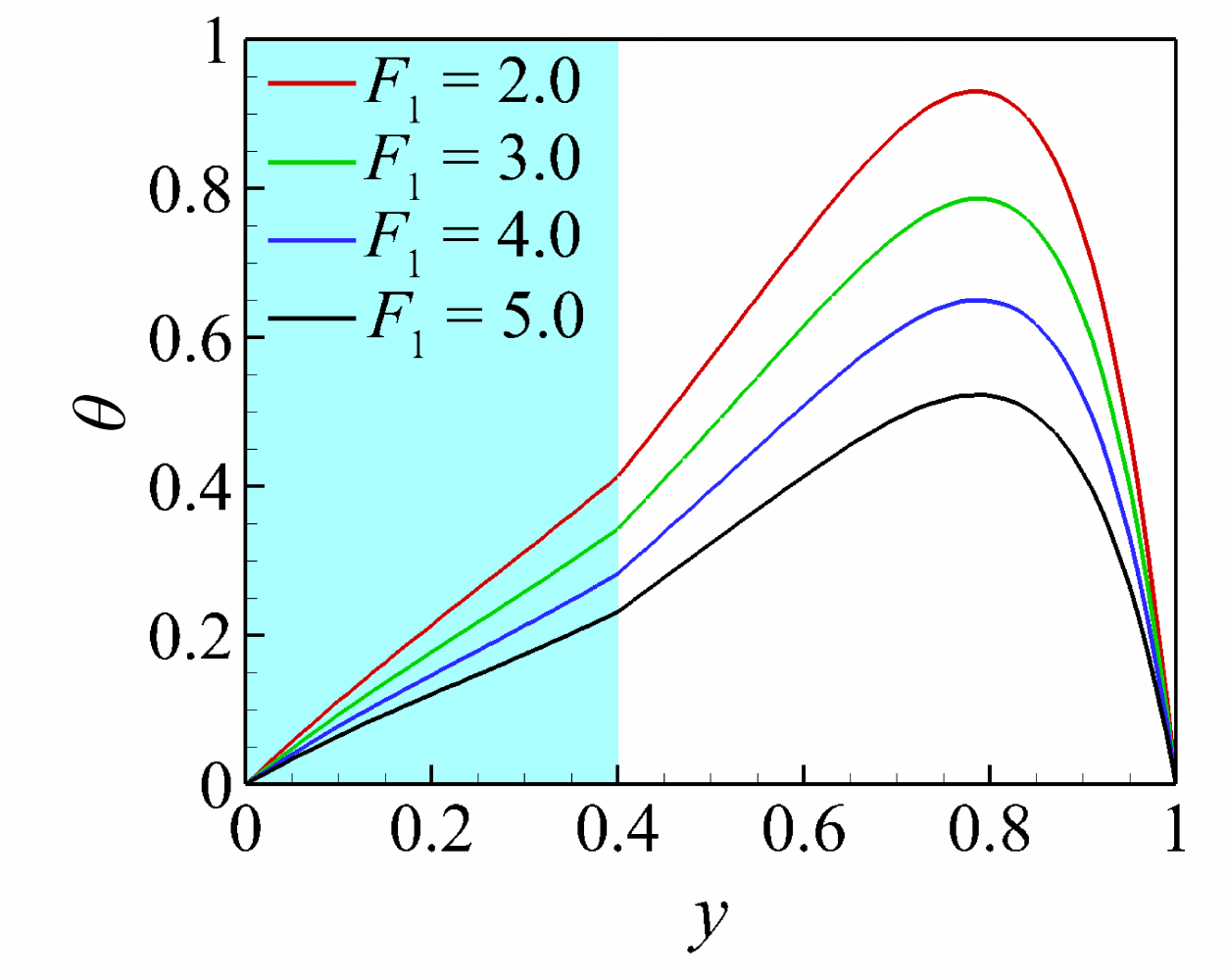}%
\caption{\label{fig14}Plot shows the temperature (\(\theta\)) distribution across the width (\(y\)) of the channel for different radiation parameters (\(F_1\)), respectively. The other parameters are, \(a=0.4\), \(\eta=2.0\), \(k=1.5\), \(\sigma=1.0\), \(\mathrm{Ha}_1=3.0\), and \(\mathrm{Br}_1=2.0\).}
\end{figure}
Figure 14 illustrates the temperature profile across the width (\(y\)) of the channel for increasing radiation parameters (\(F_1\)). Radiation parameters (\(F_i\)) signify the ratio between the radiative and conductive heat transfer. With the increase in \(F_1\), the heat depletion from the fluids inside the channel by radiation increases compared to the heat transferred due to conduction, which facilitates the sharp decrease in temperature for both the fluids inside the channel as shown in figure 14.\\
\indent Figure 15(a) shows that, in general, \(\mathrm{Be}_2\) decreases with the increase in \(F_1\). Moreover, across the width (\(0\leq{y}\leq{a}\)) of the channel, it also decreases for a particular \(F_1\). At the interface of the two fluids the magnitude of Be plummets significantly (around $\sim$ 10\% $–$ 30\%) for a particular \(F_1\). For fluid 1, the plot shows that, for a particular \(F_1\), \(\mathrm{Be}_1\) decreases initially with the increase in \(y\), goes through a minimum near \(y = 0.8\), and then increases near the upper wall. However, it can be interpreted from figure 15(a) that, other than the global minima near \(y = 0.8\), \(\mathrm{Be}_1\) always decrease with the increase in \(F_1\). Increase in radiation parameter (\(F_1\)) signifies the increase in ratio between the radiative and conductive heat transfer, which facilitates significant temperature drops in both the fluids thus ensuring the decrease in entropy generation due to the HTI (or Be). \\
\begin{figure*}
\includegraphics[width=0.85\linewidth]{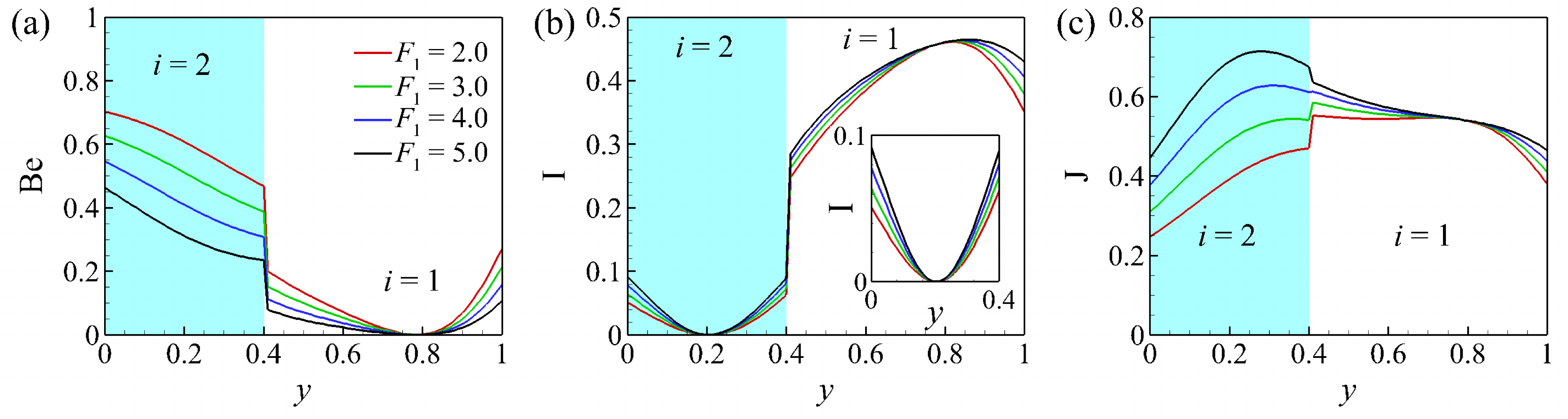}%
\caption{\label{fig15}Plots (a), (b), and (c) show the variation of Bejan number (Be), magnetic field irreversibility parameter (I) and fluid flow irreversibility parameter (J) across the width (\(y\)) of the channel for different radiation parameters (\(F_1\)), respectively. The other parameters are, \(a=0.4\), \(\eta=2.0\), \(k=1.5\), \(\sigma=1.0\), \(\mathrm{Ha}_1=3.0\), and \(\mathrm{Br}_1=2.0\).}
\end{figure*}
\begin{figure*}
\includegraphics[width=0.85\linewidth]{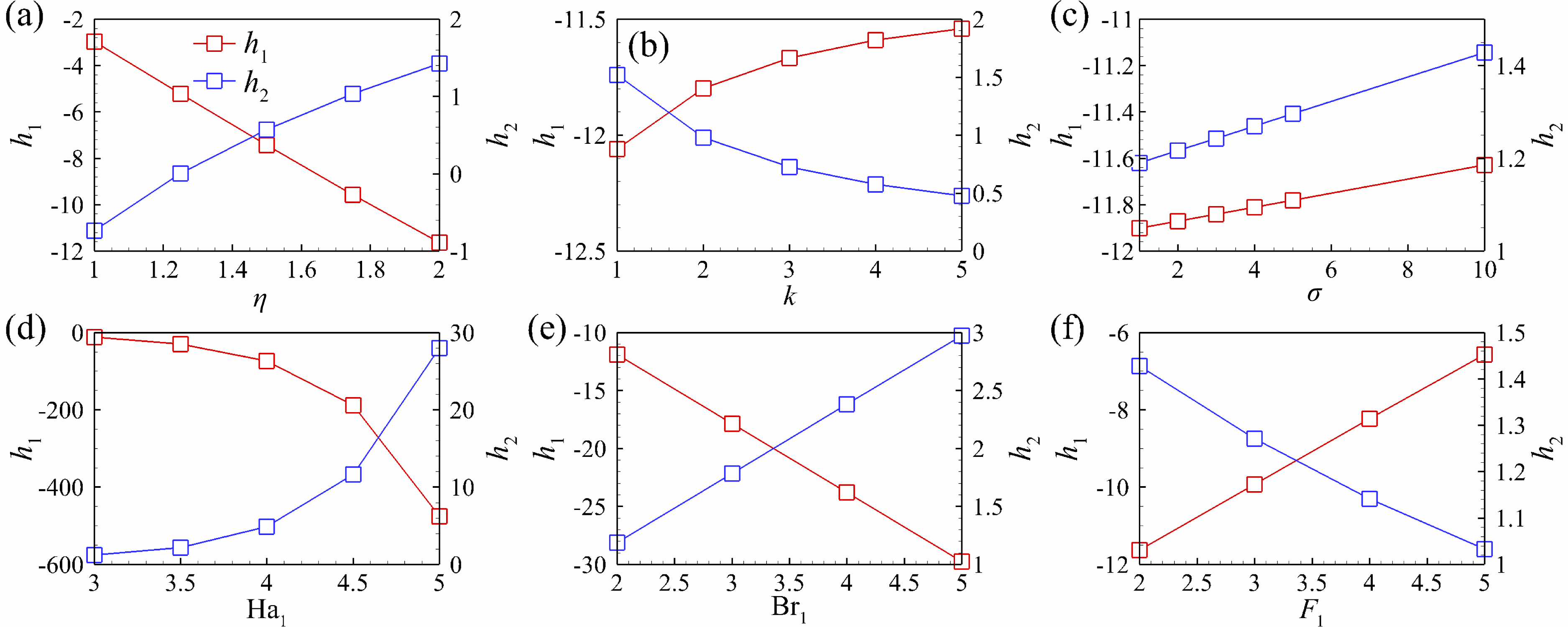}%
\caption{\label{fig16}Plots (a) $–$ (f) show the variation of  heat transfers, \(h_1\) (red symbols, left side of \(y\)-axis) and \(h_2\) (blue symbols, right side of \(y\)-axis) at the wall plates \(y = 1\) and \(y = 0\) for different viscosity ratios (\(\eta\)), thermal conductivity ratios (\(k\)), and electrical conductivity ratios (\(\sigma\)), Hartmann numbers (\(\mathrm{Ha}_1\)), Brinkman numbers (\(\mathrm{Br}_1\)), and radiation parameters (\(F_1\)), respectively. The other parameters are, \(a=0.4\) for (a) $-$ (f), \(\eta=2.0\) for (b) $-$ (f), \(k=1.5\) for (a) and (c) $-$ (f), \(\sigma=10.0\) for (a) and (f), \(\sigma=1.0\) for (b), (d) and (e), \(\mathrm{Ha}_1=3.0\) for (a) $-$ (c) and (e) $-$ (f), and \(\mathrm{Br}_1=2.0\) for (a) $-$ (d) and (f), and \(F_1=2.0\) for (a) $-$ (e).}
\end{figure*}
\indent Figure 15(b) shows that, for a particular \(F_1\),  magnetic field irreversibility parameter (I) initially decreases, goes through a minimum near \(y = 0.2\), then increases, and goes through a maximum near \(y = 0.8\). Presence of the fluid-fluid interface increases the magnitude of I by a considerably large amount (around $\sim$ 10\% $–$ 30\%). It is also interesting to note that, I is always less than 0.5 throughout the channel width for all \(F_1\), which signifies that, Be+J $>$ I, or HTI+FFI $>$ MFI. This also represents that, varying \(F_1\) essentially changes the temperature profile inside the fluids which in turn dictates the HTI and the FFI. However, plot (b) confirms that, the entropy generation due to the magnetic field (MFI) is indirectly dependent on \(F_1\). Figure 15(b) also shows that, other than the global minima (maxima) near \(y\) = 0.2 (0.8), both \(\mathrm{I}_1\) and \(\mathrm{I}_2\) always increase with the increase in \(F_1\), as shown clearly in the inset of figure 15(b).\\
\indent Figure 15(c) illustrates that, both \(\mathrm{J}_1\) and \(\mathrm{J}_2\) always increase with the increase in \(F_1\), across the channel width. It can be seen from plot (c) that, when \(F_1\) is higher (for example, \(F_1=5.0\)), the maximum contribution to entropy generation inside fluid 2 is due to FFI (contribution $\sim$ 60\%) compared to HTI (contribution $\sim$ 35\%) and MFI (contribution $\sim$ 5\%). This is because the lesser value of \(F_1\) reduces the temperatures of the fluids inside the channel which in turn decreases the entropy generation due to HTI, so that the significant portion of the entropy generation is due to the fluid flow. However, for fluid 1, the plot shows that, the changes in the magnitude of \(\mathrm{J}_1\) due to the variation of \(F_1\) are negligible across the width of the channel (\(a\leq{y}\leq{1}\)). Plot (c) also shows that irrespective of the fluids the contribution to the total entropy generation inside the channel due to FFI is always within 20\% $-$ 70\%. This signifies that, even though \(F_1\) does not necessarily change the fluid flow inside the channel as such, however, a significant portion of the total entropy generation which is due to the FFI can be controlled by tuning \(F_1\).\\
\indent In summary, figure 15 shows that, due to the change in \(F_1\), the entropy generation inside the channel can be tuned by changing the mechanism of entropy generation for both the fluids. The entropy generation due to HTI and FFI both contributes equally in the total entropy generation for fluid 2 (contribution $\sim$ 40\% each), however, in case of fluid 1 the overall entropy generation is dictated mainly by MFI ($\sim$ 30\% $–$ 50\%), and FFI ($\sim$ 50\% $–$ 60\%).\\
\indent We have also asymptotically truncated the radiative flux part from the governing equations of the present mathematical formulation to match with the results of Haddad \textit{et al.} \cite{haddad2004}. For a very low Knudsen number (Kn), the results from the present truncated formulation show a close match with the results of Haddad \textit{et al.} \cite{haddad2004}. The comparative plots of spatial variation of (a) velocity (\(u\)), temperature (\(\theta\)), and (b) local Bejan number (Be) profile are shown in figure \ref{fig18} of appendix B. 
\subsection{Rate of heat transfer}
The rate of heat transfer (\(h_1\) and \(h_2\)) per unit area at the plates \(y = 1\) and \(y = 0\) can be obtained from, \(\displaystyle \left. h_1=\frac{d\theta_1}{dy}\right|_{y=1}\) and \(\displaystyle \left. h_2=\frac{d\theta_2}{dy}\right|_{y=0}\), whereas, \(\theta_1\) and \(\theta_2\) can be evaluated using the expressions (\ref{28}) and (\ref{29}). Figures 16(a) $–$ (f) show the variation of dimensionless heat transfers, \(h_1\) (red symbols, left side of \(y\)-axis) and \(h_2\) (blue symbols, right side of \(y\)-axis) at the wall plates \(y = 1\) and \(y = 0\) for different viscosity ratio (\(\eta\)), thermal conductivity ratio (\(k\)), electrical conductivity ratio (\(\sigma\)), Hartmann number (\(\mathrm{Ha}_1\)), Brinkman number (\(\mathrm{Br}_1\)), and radiation parameter (\(F_1\)), respectively. The plots (a), (d), and (e) show that, the dimensionless heat transfer \(h_1\) (\(h_2\)) decreases (increases) with the increase in \(\eta\), \(\mathrm{Ha}_1\), and \(\mathrm{Br}_1\), respectively, whereas, the plots (b) and (f) show that, the dimensionless heat transfer \(h_1\) (\(h_2\)) increases (decreases) with the increase in \(k\) and \(F_1\), respectively. However, plot (c) shows that, with the increase in \(\sigma\), both the non-dimensional heat transfer rates (\(h_1\) and \(h_2\)) increase. The dimensionless heat transfer rates, \(h_1\) and \(h_2\), essentially depend on the slope of the temperature profile of the fluids near the lower (\(y = 0\)) and the upper (\(y = 1\)) walls.\\
\indent Figure 16 shows that, the fluid properties, like viscosity, thermal and electrical conductivity, as well as, Hartmann number (magnetic field intensity), Brinkman number (temperature gradient between the fluids and the channel wall) and radiation parameters can effectively control the rate of heat transfer from the fluids to the channel walls.

\section{Conclusions}
In summary, we identify that the radiative heat transfer and transversely applied magnetic field can significantly alter the fluid flow and the heat transfer characteristics in a two-phase non-isothermal fluid flow between two infinite horizontal parallel plates under the influence of a constant pressure gradient. The velocity profile and the temperature distribution of both the fluids can be tweaked with precision by altering the external magnetic field intensity as well as the temperature gradient between the fluids and the channel wall. Application of the transversely applied magnetic field is found to reduce the throughput and the temperature distribution of the fluids in a pressure-driven flow. Furthermore, the study also shows that, apart from the external handles, the inherent fluid properties, such as, viscosity, thermal and electrical conductivities can alter the velocity and temperature distribution significantly. In addition to that, the study confirms that the filling ratio of both the fluids can vary the velocity and temperature distribution inside the channel. The rate of heat transfer at the channel walls can also be tweaked by the magnetic field intensity, temperature gradient, and the fluid properties. \\
\indent The study also shed light on the minute details of the entropy production due to the presence of a fluid-fluid interface.  The present research addresses some of the key aspects like, the presence of two different fluids, the fluid property ratios, channel filling ratio, and the relative strength of the external fields alter the flow, heat transfer, and magnetic field irreversibility within a microchannel flow. Evaluation of the entropy generation due to the heat transfer, magnetic field, and fluid flow irreversibilities reveal that, the total entropy generation can be reduced to a minimum amount by efficiently controlling the Hartmann number, radiation parameter, Brinkmann number, filling ratio, and the fluid properties. The discontinuity in the analysis of heat transfer, magnetic field, and fluid flow irreversibilities across the interface either enhances or reduces the amount of unretractable work associated with the system. The study also predicts the amount of the entropy production due to the three different irreversibilities and their relative contribution to the total entropy generation. This information can be a key factor in designing such flow systems because the study sheds light on the optimal design parameters needed to be maintained in order to minimize or at least reduce the overall entropy production. \\
\indent The results described here can be utilized as a preliminary blue-print for developing more sophisticated MEMS devices for applications involving thermal transport. Furthermore, the present model itself can be implemented for analyzing a wide spectrum of existing EMHD transport processes across different length scales by judiciously tuning the involved non-dimensional key parameters. The entropy generation, fluid flow, and heat transfer characteristics of such a confined flow can be particularly useful in the design and fabrication of a microfluidic device which can be integrated to any state-of-the art micro emulsifiers, mixers, reactors, flow cytometers, bioanalysis tools, and drug delivery devices for improved efficiency of the existing technology. 
\begin{acknowledgments}
The author acknowledges the support from the Department of Chemical Engineering and Centre for Nanotechnology, IIT Guwahati. The author declares no conflict of interest.
\end{acknowledgments}
\appendix
\section{Mathematical Expressions}
The expressions for the coefficients \(P\) and \(Q\), in equations (\ref{28}) and (\ref{29}) of the manuscript are given by,
\begin{equation}\label{A1}
P = \frac{p_1\left(B_1+B_2a_1\right)+p_2p_3}{c_1F_1},
\end{equation}
\begin{equation}\label{A2}
Q=\frac{q_1+q_2+q_3+q_4-q_5\left(B_3+B_4b_7\right)}{\eta{M^2}d_1F_2},
\end{equation}
where,  \(p_j\) (\({j=1-3}\)), \(q_m\) (\({m=1-5}\)), \(a_n\) (\({n=1-7}\)), \(b_l\) (\({l=1-12}\)), \(c_r\) (\({r=1-8}\)), and \(d_s\) (\({s=1-11}\)) can be expressed by the following expressions (\ref{A3}) $-$ (\ref{A8}):
\begin{equation}\label{A3}
\left. \begin{array}{l}
\displaystyle
p_1=c_1F_1\left(a_2+a_3\right),\\[10pt]
\displaystyle
p_2=\mathrm{Br}_1{a_1},\hspace{2mm}\mathrm{and}\hspace{2mm}\\[10pt]
\displaystyle
p_3=c_2+c_3{a_4}+c_4{a_5}+c_5{a_6}-c_1{c_6}{a_7}.
\end{array} \right\}
\end{equation}
\begin{equation}\label{A4}
\left. \begin{array}{l}
\displaystyle
q_1=-\eta{M^4}d_2\left(b_1+b_2\right)+\eta{M^2}d_3d_6\left(b_3+b_4\right),\\[10pt]
\displaystyle
q_2=b_5d_1d_{11}\left(b_9+b_{10}\right),\\[10pt]
\displaystyle
q_3=-\eta{M^4}d_4\left(b_{10}+b_{11}\right)\left(d_5+d_6b_{12}\right),\\[10pt]
\displaystyle
q_4=b_5^2d_7d_8,\hspace{2mm}\mathrm{and}\hspace{2mm}\\[10pt]
\displaystyle
q_5=b_5b_6d_7d_9.
\end{array} \right\}
\end{equation}
\begin{equation}\label{A5}
\left. \begin{array}{l}
\displaystyle
a_1=\cosh\left(2y\sqrt{F_1}\right)+\sinh\left(2y\sqrt{F_1}\right),\\ [10pt]
\displaystyle
a_2=\cosh\bigg\{\left(\sqrt{F_1}+2\mathrm{Ha}_1\right)y\bigg\},\\ [10pt]
\displaystyle
a_3=\sinh\bigg\{\left(\sqrt{F_1}+2\mathrm{Ha}_1\right)y\bigg\},\\ [10pt]
\displaystyle
a_4=\cosh\left(3\mathrm{Ha}_1{y}\right)+\sinh\left(3\mathrm{Ha}_1{y}\right),\\ [10pt]
\displaystyle
a_5=\cosh\left(\mathrm{Ha}_1{y}\right)+\sinh\left(\mathrm{Ha}_1{y}\right),\\ [10pt]
\displaystyle
a_6=\cosh\left(4\mathrm{Ha}_1{y}\right)+\sinh\left(4\mathrm{Ha}_1{y}\right),
\\ [10pt]
\displaystyle
a_7=\cosh\left(2\mathrm{Ha}_1{y}\right)+\sinh\left(2\mathrm{Ha}_1{y}\right).
 \end{array} \right\}
\end{equation}
\begin{equation}\label{A6}
\left. \begin{array}{l}
\displaystyle
b_1=\cosh\bigg\{\left(2\sqrt{F_2}+4M\right)y\bigg\},\\ [10pt]
\displaystyle
b_2=\sinh\bigg\{\left(2\sqrt{F_2}+4M\right)y\bigg\},\\ [10pt]
\displaystyle
b_3=\cosh\bigg\{\left(2\sqrt{F_2}+M\right)y\bigg\},\\ [10pt]
\displaystyle
b_4=\sinh\bigg\{\left(2\sqrt{F_2}+M\right)y\bigg\},\\ [10pt]
\displaystyle
b_5=\cosh\left(y\sqrt{F_2}\right)+\sinh\left(y\sqrt{F_2}\right),\\ [10pt]
\displaystyle
b_6=\cosh\left(2My\right)+\sinh\left(2My\right),\\ [10pt]
\displaystyle
b_7=\cosh\left(2y\sqrt{F_2}\right)+\sinh\left(2y\sqrt{F_2}\right),\\ [10pt]
\displaystyle
b_{8}=\cosh\bigg\{\left(\sqrt{F_2}+2M\right)y\bigg\},\\ [10pt]
\displaystyle
b_{9}=\sinh\bigg\{\left(\sqrt{F_2}+2M\right)y\bigg\},\\ [10pt]
\displaystyle
b_{10}=\cosh\bigg\{\left(2\sqrt{F_2}+2M\right)y\bigg\},\\ [10pt]
\displaystyle
b_{11}=\sinh\bigg\{\left(2\sqrt{F_2}+2M\right)y\bigg\},\hspace{2mm}\mathrm{and}\hspace{2mm}\\ [10pt]
\displaystyle
b_{12}=\cosh\left(2My\right)+\sinh\left(2My\right).
\end{array} \right\}
\end{equation}
\begin{equation}\label{A7}
\left. \begin{array}{l}
\displaystyle
c_1=F_1^2-5F_1\mathrm{Ha}_1+4\mathrm{Ha}_1^4, c_2=A_2^2F_1\mathrm{Ha}_1^2c_7,\\[10pt]
\displaystyle
c_3=A_1F_1\mathrm{Ha}_1^2c_8,c_4=A_2F_1\mathrm{Ha}_1^2c_8, \\[10pt]
\displaystyle
c_5=A_1^2F_1\mathrm{Ha}_1^2c_7,c_6=2A_1A_2\mathrm{Ha}_1^2-1,\\[10pt]
\displaystyle
c_7=F_1-\mathrm{Ha}_1^2,\hspace{2mm}\mathrm{and}\hspace{2mm}c_8=F_1-4\mathrm{Ha}_1^2.
 \end{array} \right\}
\end{equation}
\begin{equation}\label{A8}
\left. \begin{array}{l}
\displaystyle
d_1=4M^4-5F_2M^2+F_2^2, d_2=A_3^2F_2\mathrm{Br}_2d_{10},\\[10pt]
\displaystyle
d_3=-A_4\mathrm{Br}_2d_9,d_4=A_3\mathrm{Br}_2d_9,\\[10pt]
\displaystyle
d_5=2A_4M^2d_{10},d_6=F_2\mathrm{Ha}_2^2,\\[10pt]
\displaystyle
d_7=-\eta{M^2}F_2d_{10},d_8=M^2A_4^2\mathrm{Br}_2,\\[10pt]
\displaystyle
d_9=4M^2-F_2,d_{10}=M^2-F_2,\hspace{2mm}\mathrm{and}\hspace{2mm}\\[10pt]
\displaystyle
d_{11}=\mathrm{Br}_2\mathrm{Ha}_2^2.
 \end{array} \right\}
\end{equation}
\\
\indent The coefficients \(B_1,B_2,B_3\) and \(B_4\) in the expressions (\ref{A1}) and (\ref{A2}) are quite cumbersome and evaluated employing the boundary conditions in the Mathematica\textsuperscript{\textregistered} code.
\section{Validation}
\begin{figure}[h]
\includegraphics[width=1\linewidth]{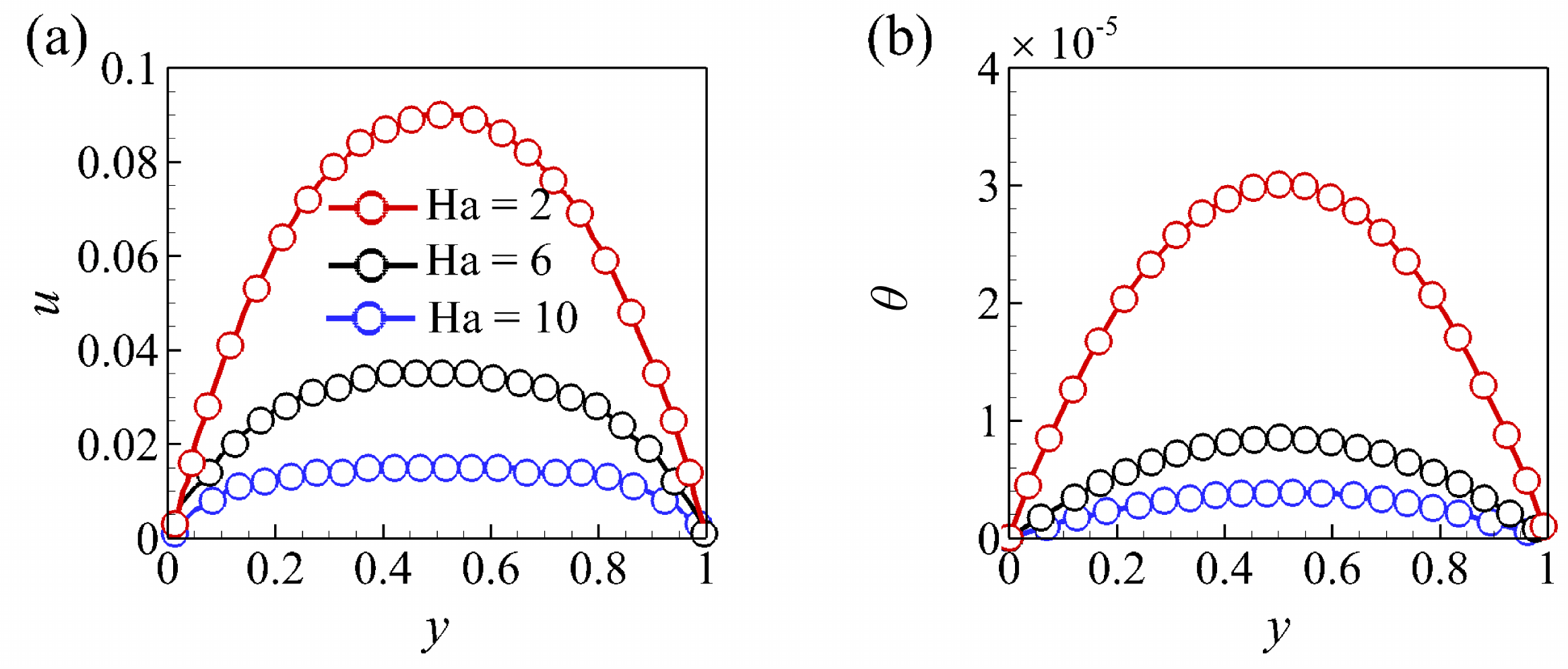}%
\caption{\label{fig17}Comparison of present study (circular symbols $–$ red, black and blue) and the analytical results (continuous lines $-$ red, black, and blue) of Duwairi \textit{et al.} \cite{duwairi2007}. Plots show the spatial variation of (a) velocity (\(u\)), and (b) temperature (\(\theta\)) profile for increasing Hartmann numbers (Ha).}
\end{figure}
\indent We have modified the governing equations mentioned in the subsections II B and II C to asymptotically match with the governing equations of Duwairi \textit{et al.} \cite{duwairi2007}. For this, we have taken the properties of both the fluids to be the same in the present framework so that it can be assumed that a single fluid is flowing through the microchannel which is the case for the case study of Duwairi \textit{et al.} \cite{duwairi2007}. Figure \ref{fig17} shows the effect of Hartmann number (Ha) on the dimensionless (a) velocity (\(u\)) and (b) temperature (\(\theta\)) profiles, respectively from the present study (discrete circular symbols $–$ red, black and blue) and the analytical (continuous lines $–$ red, black and blue) counterpart mentioned in the study of Duwairi \textit{et al.} \cite{duwairi2007}. Figure \ref{fig17} shows that with the increase in Ha, the velocity (\(u\)) and temperature (\(\theta\)) of the fluid decreases which means that the electrical conductivity dictates the velocity and the temperature profiles. These findings are similar with the present study wherein the figure \ref{fig10} shows that, with the increase in Ha both the velocity and the temperature distribution reduce significantly for a bilayer flow. \\ 
\begin{figure}[h!]
\includegraphics[width=1\linewidth]{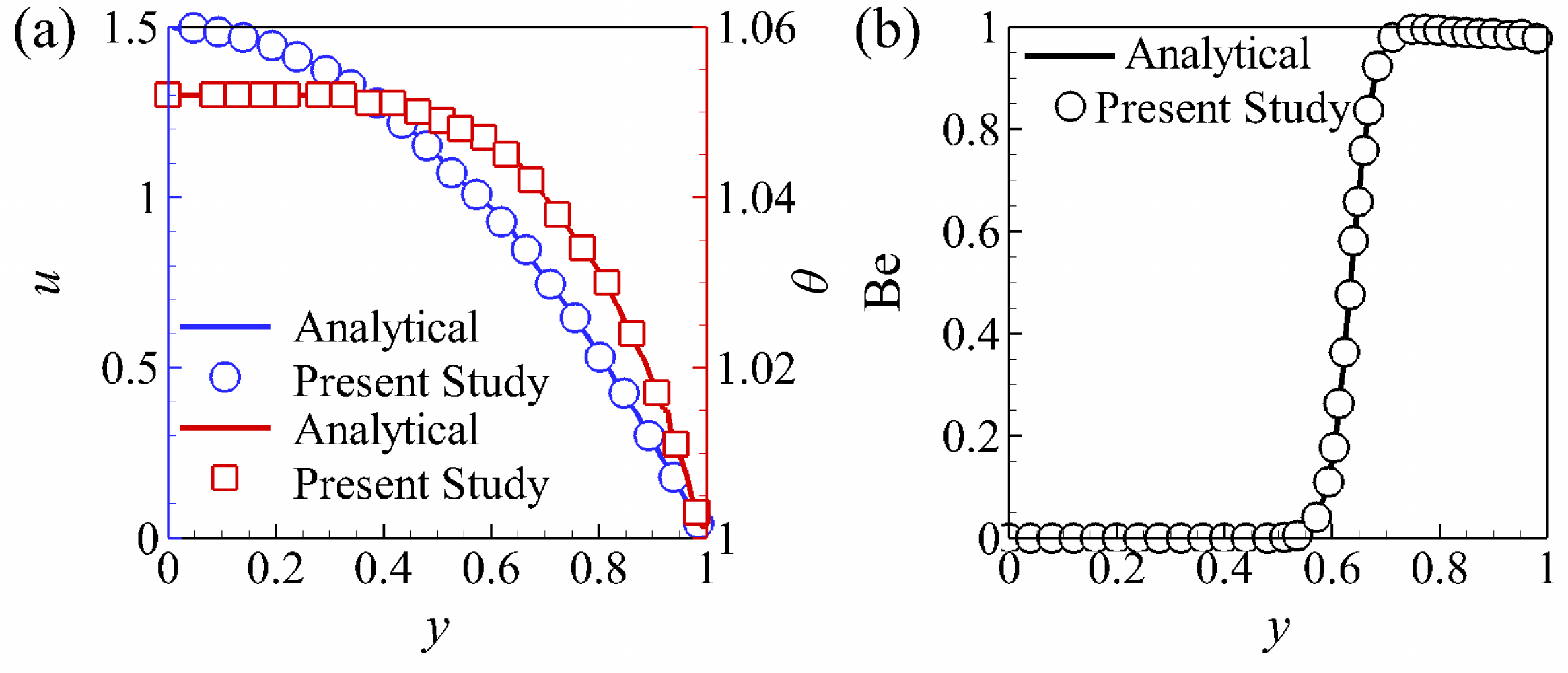}%
\caption{\label{fig18}Comparison of present study (discontinuous symbols $-$ red, blue, and black) and the analytical (continuous lines $-$ red, blue, and black) results of Haddad \textit{et al.} \cite{haddad2004}. Plots show the spatial variation of (a) velocity (\(u\) $-$ blue circular symbols and continuous line $-$ left \textit{y}-axis) and temperature (\(\theta\) $-$ red square symbols and continuous line $-$ right \textit{y}-axis), and (b) distribution of the local Bejan number (Be $-$ black circular symbol and continuous line) profile for Kn = 0.001.}
\end{figure}
\indent The set of governing equations have also been modified to match the system considered in the study of Haddad \textit{et al.} \cite{haddad2004}. Employing the same approach of the present study to non-dimensionalize the governing equations, we arrive at a similar conclusion with the previous results from analytical approach \cite{haddad2004}. Figure \ref{fig18} shows the spatial distribution of (a) velocity (\(u\)), temperature (\(\theta\)), and (b) local Bejan number (Be) profile for a very low Knudsen number (Kn = 0.001). The very low value of Knudsen number confirms that the length scale of the mean free path of the molecules are significantly less compared to the physical length scale of the system so essentially the fluid flow is in the continuum domain. The analytical results from the present study in figures \ref{fig18}(a) and \ref{fig18}(b) show a close match with the analytical results of the previous study \cite{haddad2004}. In short, figures \ref{fig17} and \ref{fig18} asymptotically validate the results from the present formulation with the results from the previous theoretical and numerical studies.

\bibliography{apssamp}
\end{document}